\documentclass[preprint,10pt]{elsarticle}

\usepackage{graphicx}
\usepackage{subfigure}
\usepackage{amssymb}
\usepackage{amsmath}

\journal{Journal of Network and Computer Applications}

\begin{document}

\def \figurename{Fig.}

\begin{frontmatter}



\title{GOSPF: an Energy Efficient Implementation of the OSPF Routing Protocol}


\author[label2]{M.~D'Arienzo}
\author[label1]{S.P.~Romano\corref{cor1}}
\ead{spromano@unina.it} 
\cortext[cor1]{Corresponding author}

\address[label1]{Department of Electric Engineering and Information Technology, University of Napoli Federico II, Italy}
\address[label2]{Dipartimento di Studi Europei e Mediterranei, Seconda Universit\`a di Napoli, Italy}

\begin{abstract}
Energy saving is currently one of the most challenging issues for the Internet research community. Indeed, the exponential growth of applications and services induces a remarkable increase in power consumption and hence calls for novel solutions which are capable to preserve energy of the infrastructures, at the same time maintaining the required Quality of Service guarantees.
In this paper we introduce a new mechanism for saving energy through intelligent switch off of network links. The mechanism has been implemented as an extension to the Open Shortest Path First routing protocol. We first show through simulations that our solution is capable to dramatically reduce energy consumption when compared to the standard OSPF implementation. We then illustrate a real-world implementation of the proposed protocol within the Quagga routing software suite.
\end{abstract}


\end{frontmatter}


\section{Introduction}
\label{sec:intro}

In this paper we present an extension of the OSPF protocol specifically conceived to optimize the overall energy consumption of the network. The algorithm we devised represents our engineering approach to the solution of the well known (though still open) issue of dynamic network topology adaptation to improve energy efficiency. The paper starts with a thorough analysis of the current achievements in the so-called \emph{green networking} field, with special regard to those approaches which have focused on the exploitation of smart energy aware routing strategies. We then highlight the major contributions of our proposal, which is based on an analytical model that takes also into account the dynamics of the traffic load the network is subject to and proposes a solution based on a fully distributed paradigm. In order to assess the performance achieved by this solution, we discuss the results of an experimental campaign based on simulations and built around an extremely realistic scenario associated with real-world traffic profiles and network topology, as inspired by the GARR Italian research network infrastructure.

The rest of the paper is structured as follows.
In section~\ref{sec:relwork} we present related works dealing with energy efficient networking, with a focus on re-engineering and dynamic adaptation approaches.
In section~\ref{sec:formulation} we provide a formal context for the definition of the problem at hand, namely the dynamic (and time-aware) adaptation of the network topology driven by energy efficiency considerations. Given the above mentioned theoretical frame, in section~\ref{sec:algorithm} we present our proposal for \emph{Green OSPF} (GOSPF), an energy efficient extension of the well-known Open Shortest Path First (OSPF) routing protocol, whose performance is thoroughly assessed in section~\ref{sec:results}. In section~\ref{sec:implementation} we move from theory to practice, by presenting a real-world implementation of the GOSPF protocol within the Quagga routing protocol suite. Finally, section~\ref{sec:conclusions} summarizes the main results of our current efforts and indicates the most interesting directions of our future work.

\section{Related Work}
\label{sec:relwork}

The power saving issue in the ICT era has attracted more and more attention
in the last few years.
In~\cite{GUP03}, a pioneering analysis about the impact of energy
consumption on network devices in the Internet has been presented.
The authors argue the need to change the network protocols in order to save energy by putting unused interfaces and devices in a sleep state.
The main approaches to reduce energy consumption in networking infrastructures usually focus on three different areas: the network edge components (including PCs and other end-devices), the access networks, and the backbone. To reduce the power consumption in any of these areas, several novel solutions have been proposed. Most of these solutions are based on three fundamental approaches: re-engineering, dynamic adaptation, and sleeping/standby \cite{BOL11}.

Re-engineering is concerned with the design of innovative technologies that claim more efficiency in the network equipment. It includes the optimization of hardware components as well as the implementation of innovative network paradigms that can contribute towards a reduction in power consumption. The dynamic adaptation of the network resources, instead, provides mechanisms for tuning packet processing engines and network interfaces in order to satisfy at the same time traffic load and performance requirements, while lowering the demand for energy. Finally the sleeping/standby approach envisages switching either unused or underutilized network components to a low power consumption mode. Due to the requirement to keep devices (specifically the end-systems) always connected to the network, this approach calls for proxying techniques capable to guarantee the ``presence'' of the component also when it is in sleeping or standby mode.
Actually, the selection of a specific approach mainly depends on the particular network scenario upon which the mechanism will be implemented.

With regard to PCs and end-devices, the common approach is to implement a sleeping mechanism. In~\cite{CHR08} the authors propose to turn into sleeping mode the network interfaces of the devices when no data are going to be transmitted. As mentioned above, since several network protocols require that devices are continuously connected, a ``proxy'' component, called Network Connectivity Proxy (NCP), keeps on guaranteeing the presence of a network interface even when it switches to sleeping mode. This mechanism requires a specific network protocol to wake up the interface when data are sent to the end-device. Similarly, in~\cite{SAB08} a connection proxy mechanism is directly implemented in the network interface card (NIC). Such a solution allows to power off the end-device while keeping network connectivity.

Solutions adopted for energy efficiency in the access network mainly exploit both the re-engineering and the dynamic adaptation approaches. As an example, in~\cite{PAR10} the authors propose a mechanism to reduce the power consumed by a Home Gateway (HG). They implement the HG through several functional blocks which can be powered down to reduce energy, with the exception of the so-called NPA (Network Protocol Agent), namely the component which is in charge of maintaining network connectivity.

Re-engineering and dynamic adaptation are also exploited in the network backbone. In this context a valuable work has been proposed by Baldi and Ofek in~\cite{BAL09}. The authors propose a re-engineering of the network infrastructure through a \emph{pipeline forwarding} of IP packets aimed at optimizing the overall energy consumption of the single network elements. By synchronizing all routers operations in the network with the same clock reference, it is possible to optimize packet forwarding for periodic traffic, such as video. The proposed approach reduces per packet processing, limits the memory requirements, and allows a full link utilization. In support of the use of these approaches, ~\cite{CHA08} analyzes the power consumption of the routers in several configuration scenarios. The results of this study confirm the need for a re-engineering of the network equipment that should support power-awareness mechanisms to considerably reduce energy consumption.
With regard to dynamic adaptation, work in~\cite{NED08} considers two fundamental schemes for power management: sleeping and rate adaptation. The former puts network interfaces to sleep for a short idle period. This solution is supported by a smart buffering mechanism at interface level in order to avoid packet losses during sleeping time and to aggregate traffic in bursts. The second mechanism allows to adapt the transmission rate of a device based on the sensed traffic conditions. Actually, the rate should not considerably impact QoS performance figures. ~\cite{BOL09} still relies on improved efficiency of network equipment by adopting a modular architecture. The work proposes an optimization policy based on an analytical framework able to reduce power consumption of a network device with respect to its expected forwarding performance.

Several works also address the reduction of backbone energy consumption through an adaptation of the network topology according to different conditions of network traffic. These approaches try to adapt network capacity in terms of links and routers. They limit the number of network elements active and reroute all the traffic through a minimum number of nodes and links. In~\cite{CHI09a} the authors propose an algorithm to define a minimum set of nodes and links capable to support a given traffic. The approach is then applied to an actual ISP network topology in~\cite{CHI09b}.
Similarly, in~\cite{TUC08} the energy consumption is optimized by dynamically shutting off portions of the network and rerouting traffic by meeting demanded traffic profiles.
In~\cite{CUO11} an algorithm based on \emph{algebraic connectivity} parameters is proposed to identify links to be powered off while preserving the overall network connectivity. However, none of these proposals includes time-dependency in its evaluation.

Additional solutions rely on the insertion of energy awareness in routing algorithms. The work presented in ~\cite{RES09} introduces several equipment energy profiles defined as a function of the energy consumption and the network traffic load. These profiles are then kept into account by the routing algorithm to help route traffic in a way that reduces the overall energy consumption of the network. Not differently, ~\cite{CIA11} suggests an energy aware enhancement of the well-known routing protocol OSPF (Open Shortest Path First). Rather than computing a Shortest Path Tree (SPT) for each router in the network, the authors introduce the concept of ``SPT exportation'', where a set of routers, called ``exporters'', forces the use of their SPTs to other routers, called ``importers''. The algorithm can be implemented in a centralized way by limiting the impact on the OSPF.

Our solution falls in the group of dynamic adaptation mechanisms for the network backbone. In particular, a network topology adaptation approach is exploited to identify the links that can be switched off in order to optimize the overall network resources. Differently from~\cite{CHI09a} and~\cite{CHI09b} our solution is based on an analytical model that also takes into account the dynamics of traffic load. Similarly to~\cite{CIA11} we consider an enhancement of the OSPF protocol, but in our case a completely distributed and integrated solution is developed. Also, connectivity parameters are interpreted in a way that is similar to the solution proposed in ~\cite{CUO11}, but it is now extended to the time-dependent case.

\section{Problem Formulation}
\label{sec:formulation}
 
The problem of dynamic network topology adaptation for energy efficiency can be summarized as follows. Given a physical network topology and the instantaneous traffic demanded, the challenge is to find the minimum number of links which both reduces the power consumption and accommodates traffic requests without exceeding the maximum capacity of the remaining links.

In order to correctly analyze this problem an Integer Linear Programming (ILP) formulation has been provided. Let $G = (V,E)$ be a graph with the vertex set $V$ and edge set $E$. Such a graph represents a network topology with $N = |V|$ nodes and $L = |E|$ links. Let $u_{ij}$ be the maximum capacity of the link
$(i,j)$ between node $i$ and node $j$, and let $\alpha \in [0,1]$ be
the maximum link utilization tolerated. Let $W=\{w_{sd}\}$ be the
set of traffic demands, where $w_{sd}$ is the amount of traffic
going from the source $s$ to the destination $d$, with
$s,d=1...N$. Furthermore, we consider $c_{ij}$ to be the routing cost related to the link $(i,j)$.

Let $x_{ij} \in \{0,1\}$ be an integer variable that is $1$ when
the link $(i,j)$ is powered on and $0$ otherwise, and let $y_{ij}^{sd}$ be the distribution decision variable indicating
the amount of demanded traffic between $s$ and $d$ allocated on
link $(i,j)$.

Finally, let $P_{ij}$ be the power consumed by link $(i,j)$ when it is switched on. Based on the previous assumptions, the optimization problem described above can be formulated as follows:

\begin{equation}
    min \sum_{i,j=1}^{N} P_{ij} x_{ij} + \sum_{s,d=1}^{N}\sum_{i,j=1}^{N} c_{ij} y_{ij}^{sd}
    \label{equ:cmndproblem}
\end{equation}

subject to

\begin{equation}
    \sum_{j=1}^{N} y_{ij}^{sd} - \sum_{j=1}^{N} y_{ji}^{sd} =
    \begin{cases}
        w_{ij}^{sd} & \forall s,d,i = s \\
        - w_{ij}^{sd} & \forall s,d,i = d \\
        0           & \forall s,d,i \neq s,d \\
    \end{cases}
    \label{equ:flowconservation}
\end{equation}

\begin{equation}
    \sum_{s,d=1}^{N} y_{ij}^{sd} \leq \alpha u_{ij} x_{ij} \;\;\; \forall i,j
    \label{equ:linkcapacity_I}
\end{equation}

\begin{equation}
    y_{ij}^{sd} \in [0,w_{ij}^{sd}] \;\;\; \forall i,j,s,d
    \label{equ:flowdistrconstraint}
\end{equation}

\begin{equation}
    x_{ij} \in \{0,1\} \;\;\; \forall i,j
    \label{equ:linkdistrconstraint}
\end{equation}

The formulation represents a typical \emph{Capacitated Multicommodity
Network Design} (CMND) problem, widely adopted in transportation,
logistics and telecommunications. Given a network with limited
capacity, the final objective of such a problem is to select the
links to include in the final version of the network in order to
minimize the total system cost, computed as the sum of a cost due
to the link utilization and a routing cost, while the demanded commodities
are transported by the network.

In our formulation the cost due to the link utilization is the power
consumed by the link. In the objective function
(\ref{equ:cmndproblem}) the first term represents the total power
consumed by the network with a specific link configuration
$\{x_{ij}\}$, while the second term represents the total cost for
routing the demanded traffic along specific paths. Equation (\ref{equ:flowconservation}) represents the flow
conservation constraint, while equation
(\ref{equ:linkcapacity_I}) limits the total traffic allocated on an
active link at $\alpha$ times its maximum capacity.

In our work we consider that the traffic from source
to destination can be routed on a single path according to the
most common IGPs (Interior Gateway Protocols). For this reason we
assume that the amount of traffic $y_{ij}^{sd}$ from node $s$ to
node $d$ allocated on link $(i,j)$ can be either $0$ or $w_{ij}^{sd}$
depending on whether or not link $(i,j)$ belongs to the selected path between
source and destination. Clearly, from equation
(\ref{equ:linkcapacity_I}), $y_{ij}^{sd} = 0$ when $x_{ij} = 0$.

A similar formulation has already been adopted in~\cite{CHI09a}.
Unfortunately this model neglects the time-dependent properties
of the demanded traffic. In a real application, indeed, the
traffic pattern can vary significantly over time. By using the
average value of the traffic only a suboptimal optimization can be
achieved. A proper configuration of the network in accordance with
demanded traffic should be provided in order to increase
the saved energy. Furthermore the network could be unable to
support unexpected peaks of traffic, hence increasing the possibility of
losing packets.

Based on the above considerations, a time-dependent reformulation
of the problem should be provided in order to also take into account the dynamics of traffic generation in the optimization process. Let us assume that the graph $G=(V,E,T)$ represents a network with $|V|$ nodes, $|E|$
links, and a time horizon $T$. We suppose that an instantaneous
traffic $w_{sd}(t)$ is demanded by each pair of source and
destination nodes $(s,d)$ in the interval $[0,T]$.

The time-continuous formulation of the problem becomes

\begin{equation}
    min \sum_{i,j=1}^{N} P_{ij} \int_{t_{1}}^{t_{2}} x_{ij}(t) \mathrm{d}t + \sum_{s,d=1}^{N}\sum_{i,j=1}^{N} c_{ij} \int_{t_{1}}^{t_{2}} y_{ij}^{sd}(t) \mathrm{d}t
    \label{equ:cmndproblem-continuous}
\end{equation}

$\forall t_{1},t_{2} \in [0,T]$, subject to

\begin{equation}
    \sum_{j=1}^{N} \int_{t_{1}}^{t_{2}} y_{ij}^{sd}(t) \mathrm{d}t - \sum_{j=1}^{N} \int_{t_{1}}^{t_{2}} y_{ji}^{sd}(t) \mathrm{d}t = \hat{w}_{ij}^{sd}
    \label{equ:flowconservation-continuous}
\end{equation}

where

\begin{equation}
    \hat{w}_{ij}^{sd} =
    \begin{cases}
        \int_{t_{1}}^{t_{2}} w_{ij}^{sd}(t) \mathrm{d}t & \forall s,d,i = s \;\;\; \forall t_{1},t_{2} \in [0,T] \\
        - \int_{t_{1}}^{t_{2}} w_{ij}^{sd}(t) \mathrm{d}t & \forall s,d,i = d \;\;\; \forall t_{1},t_{2} \in [0,T]\\
        0           & \forall s,d,i \neq s,d \;\;\; \forall t_{1},t_{2} \in [0,T] \\
    \end{cases}
    \label{equ:flowconservation2-continuous}
\end{equation}

\begin{equation}
    \sum_{s=1}^{N}\sum_{d=1}^{N} y_{ij}^{sd}(t) \leq \alpha u_{ij} x_{ij}(t) \;\;\; \forall i,j \;\;\; \forall t \in [0,T]
    \label{equ:linkcapacity}
\end{equation}

\begin{equation}
    y_{ij}^{sd}(t) \in [0,w_{ij}^{sd}(t)] \;\;\; \forall i,j,s,d  \;\;\; \forall t \in [0,T]
    \label{equ:flowdistrconstraint-continuous}
\end{equation}

\begin{equation}
    x_{ij}(t) \in \{0,1\} \;\;\; \forall i,j \;\;\; \forall t \in [0,T]
    \label{equ:linkdistrconstraint-continuous}
\end{equation}

The discrete-time formulation of the problem derives from the
continuous-time model by replacing in equations
(\ref{equ:cmndproblem-continuous})--(\ref{equ:linkdistrconstraint-continuous})
$\int_{t_{1}}^{t_{2}}$ with $\sum_{t_{1}}^{t_{2}}$ and $\forall
t_{1},t_{2} \in [0,T]$ with $\forall t_{1},t_{2} \in
K=\{1,2,...T-1\}$.

The problem of ``flow over time'' was first introduced by Ford and
Fulkerson~\cite{FOR58}. It usually concerns the problem of maximum
flow in the presence of a finite link traversal time. Recently, this
problem has been also extended to \emph{Dynamic Generative Network Flow}
(DGNF) problems where, in spite of the previous formulation, flows
are generated and absorbed at source and destination dynamically
over time, while transmission time is
instantaneous~\cite{FAT10}. A common approach to solve
time-dependent network flow problems, including the CMND problem,
is to transform them into static problems by using an
auxiliary time expanded network
$G^{T}(V^{T},E^{T})$~\cite{FOR58}~\cite{FLE02}~\cite{MUR04}. Such an
approach requires a discrete-time formulation of the problem for every period of length $T$. Given a dynamic graph $G(V,E,T)$ the time-expanded
network $G^{T}(V^{T},E^{T})$ is defined as follows:

\begin{eqnarray}
    V^{T}=\{v^{t}|v \in V,t \in K\} \\
    E^{T}=\{(i,j)^{t}|(i,j) \in E,t \in K\} \\
    {w_{sd}}^{t}=w_{sd}(t) \;\;\; \text{for} \;\;\; s^{t},d^{t} \in V^{T}
\end{eqnarray}

The graph $G^T$ represents a network with $|V^{T}|=N \cdot T$
nodes and $|E^{T}|=L \cdot T$ links. Given a source $s$ and a destination $d$ in the graph $G$, the time-expanded graph $G^T$ contains $T$ sources $s^0,s^1,...s^{T-1}$ and $T$ destinations $d^0,d^1,...d^{T-1}$. Our optimization problem can actually be reformulated as follows:

\begin{equation}
    min \sum_{t \in K}\left( \sum_{i,j=1}^{N} P_{ij} x_{ij}^{t} + \sum_{s,d=1}^{N}\sum_{i,j=1}^{N} c_{ij}
    {y_{ij}^{sd}}^{t}\right)
    \label{equ:cmndproblem-expanded}
\end{equation}

subject to

\begin{equation}
    \sum_{j=1}^{N} {y_{ij}^{sd}}^{t} - \sum_{j=1}^{N} {y_{ji}^{sd}}^{t} =
    \begin{cases}
        {w_{ij}^{sd}}^{t} & \forall s^t,d^t,i = s^t \\
        - {w_{ij}^{sd}}^{t} & \forall s^t,d^t,i = d^t \;\;\;\;\; t \in K\\
        0           & \forall s^t,d^t,i \neq s^t,d^t \\
    \end{cases}
    \label{equ:flowconservation-expanded}
\end{equation}

\begin{equation}
    \sum_{s,d=1}^{N} {y_{ij}^{sd}}^{t} \leq \alpha u_{ij} x_{ij}^{t} \;\;\; \forall (i,j)^t,s^t,d^t \;\;\; t \in K
    \label{equ:linkcapacity-expanded}
\end{equation}

\begin{equation}
    {y_{ij}^{sd}}^{t} \in [0,{w_{ij}^{sd}}^{t}] \;\;\; \forall (i,j)^t,s^t,d^t \;\;\; t \in K
    \label{equ:flowdistrconstraint-expanded}
\end{equation}

\begin{equation}
    x_{ij}^{t} \in \{0,1\} \;\;\; \forall (i,j)^t \;\;\; t \in K
    \label{equ:linkdistrconstraint-expanded}
\end{equation}

with

\begin{eqnarray}
    x_{ij}^{t}=x_{ij}(t) \;\;\; \text{for} \;\;\; (i,j)^{t} \in     E^{T}\\
    {y_{ij}^{sd}}^{t}=y_{ij}^{sd}(t) \;\;\; \text{for} \;\;\; (i,j)^{t} \in E^{T} \;\; s^{t},d^{t} \in V^{T}
\end{eqnarray}

Although the time expanded network simplifies the resolution of a dynamic CMND problem, it is still an NP-hard problem and exact methods can be adopted just for solving trivial cases. In this paper we propose a heuristic for the dynamic optimization of network resources according to demanded traffic.

From equation (\ref{equ:cmndproblem-expanded}) the following assumptions for the algorithm design can be made: (i) for each discrete time $t$ a suboptimal solution which accommodates the requested traffic \{$w_{sd}(t)$\} can be identified; (ii) no optimization is needed in the presence of negligible changes in the demanded traffic (i.e.~$w_{sd}(t_1) \approx w_{sd}(t_2)$ for $t_1 \leq t_2$); (iii) only significant modifications in the traffic dynamics should require a new network design optimization; (iv) a suboptimal solution can still be identified if we separately consider the optimization of the total power consumed by the network (i.e.~the first term in the sum) and the routing cost (i.e.~the second term in the sum).

The proposed algorithm includes all the mentioned features. In particular, it  extends a well known routing algorithm, the OSPF (Open Shortest Path First), with a mechanism for reducing the number of active links. In this way we can optimize both the power consumption (through our new mechanism) and the routing cost (through the OSPF itself). Furthermore, activation and deactivation of the links, as well as standard routing operations, are only triggered when the traffic changes significantly. In the following section we will detail our solution.

\section{A ``Green'' OSPF}
\label{sec:algorithm}

Our basic idea concerns the definition of an algorithm capable of adapting the network topology to the traffic profile to be supported, while minimizing the overall network utilization level. Indeed, the traffic can significantly change during the day, so an optimal configuration can only be achieved through dynamic adaptation of the network resources. Variations in the traffic profile should actually trigger a new configuration of the network topology. Traffic increases require a change in the network configuration in order to be effectively put through. Similarly, when the traffic decreases, a more efficient configuration of the active links should be considered. The new topology is then used by a routing algorithm to optimize routing costs.

From a formal point of view, the optimal graph $G'$ should vary over time between the full graph and a minimum graph as follows:

\begin{equation}
    G_{min}(V,E_{min}) \subseteq G'(t)(V,E'(t)) \subseteq G(V,E)
\end{equation}

where

\begin{equation}
    E_{min} \subseteq E'(t) \subseteq E
\end{equation}

A minimum graph is introduced in order to guarantee network connectivity at all times.

Topology adaptation is realized through link cutting and link grafting mechanisms. The main idea is to cut only underloaded links: we assume that the traffic transmitted across such links can be supported by other links in the network. To this purpose, two important issues have to be taken into account: (i) no link cutting should disconnect the network; (ii) cutting should not involve the links with high capacity. Every node in the network must in fact be reachable all the time. For the second issue, the system should start to cut the links with lower capacity values in order to allow the reallocation of the traffic on other links. Whenever a link becomes overloaded, instead, link grafting takes place: new links are iteratively added to the topology until the network becomes capable to support the required traffic demand. A proper mechanism for the selection of the links to be included in the topology has been realized.

Actually, topology adaptation does not optimize also the routing cost. In our solution this aspect is delegated to the well-known OSPF routing algorithm. Such an approach, which reflects the idea to separate topology from routing optimization, allows: (i) to neglect all the issues related to routing by exploiting an existing (and effective) routing protocol; (ii) to propose a feasible solution for energy efficiency by just slightly modifying a widely adopted protocol; (iii) to realize a completely distributed topology management scheme by exploiting the advertisement mechanism of the underlying routing protocol.

Therefore, every time a traffic variation requires a topology change the control messages of OSPF are used to inform all nodes about the need to redefine the network topology. The new network topology is then used by the Dijkstra's algorithm to compute the new paths.

In the following of this section we will first provide some background information about standard operation of the OSPF protocol; then, we will delve into the details of our proposal for an energy-efficient version of OSPF.

\subsection{The OSPF protocol: a quick overview}

Open Shortest Path First (OSPF) is a routing protocol for IP-based networks. It falls into the category of the so-called interior routing protocols, since it operates within the boundaries of a single autonomous system (AS). It relies on a link state routing algorithm through which it gathers link state information from available routers and constructs a topology map of the network. Such a map is referred to as the \emph{Link State Database} (LSDB). Each participating router has an identical database and each individual piece of this database is a particular router's local state, namely the router's usable interfaces and reachable neighbors. The router distributes its local state throughout the Autonomous System by flooding ad hoc messaged known as \emph{Link State Advertisements} (LSA). All routers run the exact same algorithm, in parallel.  Starting from the constructed LSDB, each router builds a tree of shortest paths with itself as root. This shortest-path tree is based on Dijkstra's shortest path algorithm and gives the route to each destination in the Autonomous System.  The OSPF routing policies for constructing a route table are governed by link cost factors associated with each routing interface. Cost factors are typically expressed as simple unit-less numbers and may depend on the distance (e.g., round-trip time) of a router, on data throughput of a link, or on properties like link availability and reliability. 

\subsection{Designing the G-OSPF protocol}

First of all we suppose that for every interface of the network nodes three operational states exist: (i) \emph{active}, when the interface processes packets, (ii) \emph{idle}, when the interface is powered on but no packets are processed, and (iii) \emph{sleep}, when the interface is in low energy configuration and no data are elaborated.

Given an interval $T$, the interface can switch among the three states respectively for an activation time $T_{ac}$, an idle time $T_{id}$, and a sleeping time $T_{sl}$.

Based on this assumption we adopted the following energy model for the network interface $i$:

\begin{equation}
\label{eq:energy}
    E_{i}=P_{a}*T_{ac} + P_{i}*T_{id} + P_{s}*T_{sl}+E_{c}*C
\end{equation}

where $P_{a}$ is the power consumed by the interface in active state and is a function of the managed traffic, $P_{i}$ is the power consumed in idle state, $P_{s}$ is the power consumed in sleeping mode, $E_{c}$ is the energy consumed to make the transition from a sleep state to an idle state, and finally $C$ is a counter that keeps track of the number of switches.

In our algorithm every node computes a \emph{Maximum Capacity Spanning Tree} (MCST) through the well-known Kruskal's algorithm~\cite{KRU56}. Such an algorithm actually allows for the computation of a \emph{Minimum Spanning Tree} (MST). In our case, the MST becomes a MCST due to the fact that we assign each link a cost that is proportional to the inverse of its capacity (i.e., the higher the capacity of the link, the lower its cost).

The MCST allows to minimize the number of links (equal to $|V|-1$), to maximize the network capacity, and to keep the network always connected. Therefore it guarantees the reachability of all the network nodes, as well as the allocation of the maximum traffic load on the minimum number of links.

During the selection of the links to be put in sleep mode, only the edges that don't belong to the MCST can be cut off. This guarantees both the connectivity of the network and the possibility to switch off the highest number of links. Thanks to the global view of the network offered by the OSPF protocol, the MCST computed at the single node is the same as the one computed at any other node in the network.

The set of nodes in the topology which have to be configured in sleep mode is then computed.
A link is active if both the interfaces it connects are active. In order to disable a link both its edges must be disabled. The utilization rate of the interfaces is the key feature we rely upon in order to decide whether or not to switch network nodes from an active state to a sleep state (and viceversa). Such an indicator is computed as the ratio between the activity time $ T_{ac}$ and the total time $T$.

\begin{center}
\textbf{$ U_{r}=\frac{T_{ac}}{T} $}
\end{center}
\begin{center}
$T_{ac}=\frac{b_{T}}{br}$
\end{center}

where $b_{T}$ is the total number of elaborated (i.e., either transmitted or received) bits, and $br$ is the traffic rate (both incoming and outgoing) expressed in bits per second.

Each node periodically controls the percentage of utilization of its interfaces. Two thresholds have been defined in order to evaluate interface utilization: an \emph{upper threshold}, that we call $\gamma _{u}$, and a \emph{lower threshold}, indicated with $\gamma _{l}$.

These two values indicate the interval in which the utilization rate $U_{r}$ must be included in order to be considered acceptable. If $U_{r} > \gamma_{u}$ then the interface is overutilized and the node starts the \emph{link grafting process}. On the other hand, if $U_{r} < \gamma_{l}$ the interface is underutilized and, provided that no other interface is overutilized, the node starts the \emph{link cutting process}.

During the link cutting process all nodes perform the following actions:

\begin{itemize}
	\item they switch to sleep state all interfaces if they do not belong to the MCST;
	\item they notify such an event to the other routers in the network.
\end{itemize}

The other interface connected to the link in question is also switched off as soon as the peering node receives the related notification.
The notification is done by extending the OSPF advertisement messages with a new \emph{Link State Cut Update Packet} (LSCUP) that is sent to the other nodes in the network, and which contains the ID of the link to be put in sleep mode.

Each node receiving the LSCUP operates as follows:

\begin{itemize}
	\item if the link is associated with one of its interfaces, the interface in question is put into sleep state;
	\item the LSCUP is broadcast to the other nodes.
\end{itemize}

\begin{figure}
  \centering
  \includegraphics[width=0.65\textwidth]{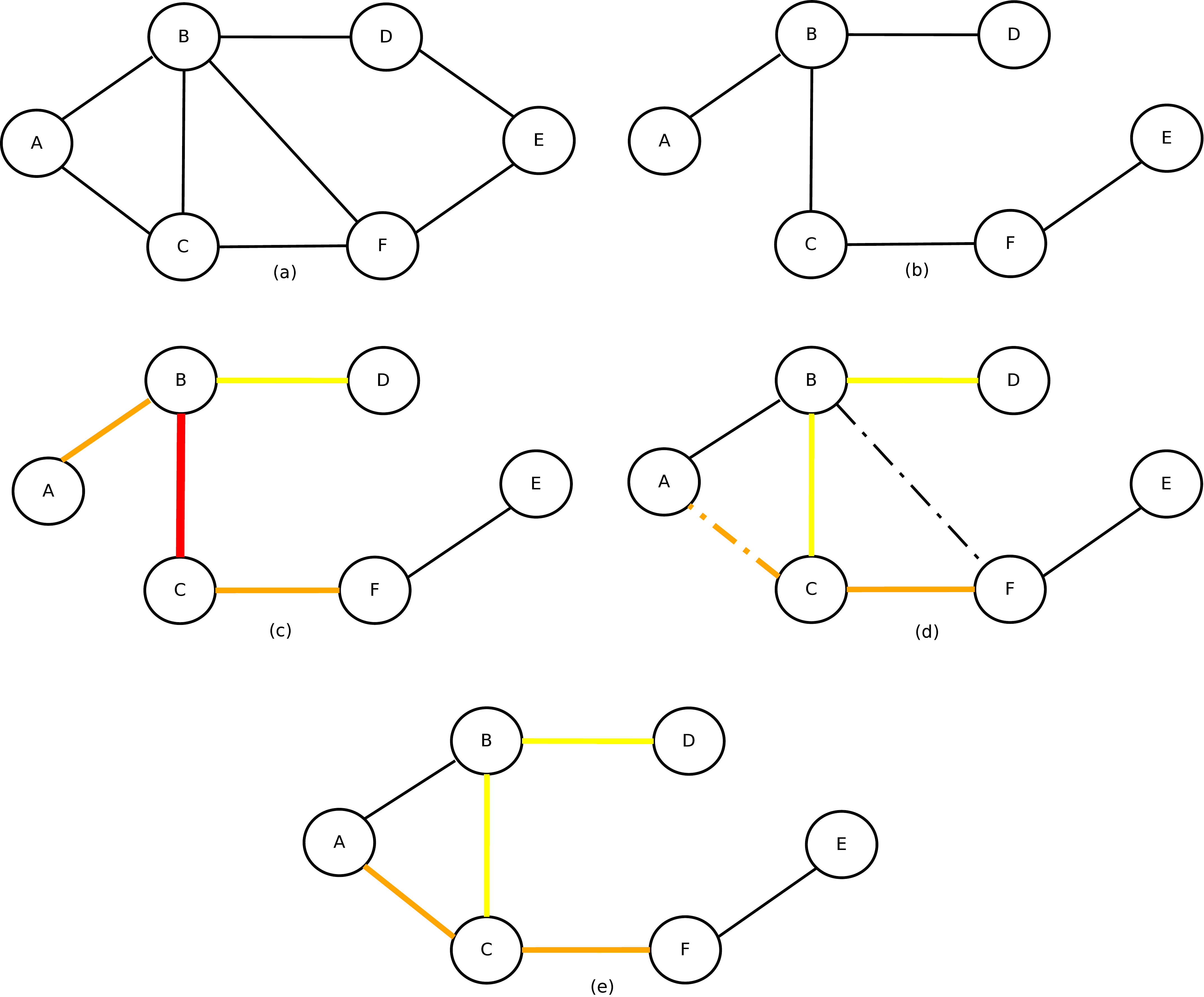}
  \caption{Graft Algorithm}\label{fig:graft}
\end{figure}

In the link grafting process, instead, when a node detects an overload situation on one of its interfaces, it operates as follows:

\begin{enumerate}
	\item it notifies the overload to the other nodes in the network;
	\item it restores the available links.
\end{enumerate}

The link to restore is identified based on a matrix that each node maintains and which contains the links that have been previously switched off. The edges which are closest to the overloaded link are hence restored. In particular, all the links at $0$ hop distance from the congested node are first activated and a notification is sent to the other nodes. If the link is still congested, then all the links at $1$ hop distance are switched on, and so on until no more congested link is present in the network.

The mentioned operations are carried out by every node receiving the notification, but only those nodes which are connected to restored links physically switch on their interfaces.

To avoid overheads, as well as oscillations of the active links, a restored link cannot be switched off again for the entire duration of a safeguarding time interval. The notification is done by sending a \emph{Link State Graft Update Packet} (LSGUP), which contains information about the links to be restored.
The node that receives the LSGUP forwards the packet to its peering nodes, and starts the link restoring process, if needed.

In \figurename\ref{fig:graft} an example of the grafting process is reported. In particular, let us consider the original network topology associated with the graph in \figurename\ref{fig:graft}(a).  If no initial traffic flows across such a network, the \emph{Cut} phase cuts all links that don't belong to the MCST that is shown in \figurename\ref{fig:graft}(b). After a while, two traffic flows are transmitted: the former from node $D$ to node $C$, the latter from node $F$ to node $A$. We also make the hypothesis that the aggregated throughput of these two flows causes an overload on link $B-C$ (\figurename\ref{fig:graft}(c)). This forces either node B or node C  (or even both) to start the \emph{Graft} phase. The node in question restores all the links belonging to the above mentioned switched-off links matrix (those in the first not empty row), as we can see in \figurename\ref{fig:graft}(d). Finally, since the link B-F is not used, either B or F cuts it. Eventually, the updated graph will look like the one in~\figurename\ref{fig:graft}(e). Link $C-A$ has been added to the original tree and it will be used in order to carry the traffic originating in $F$ and addressed to $A$ (along the newly created path $F-C-A$). With this new configuration, the link $B-C$ is not overloaded anymore.

%
%

%
\section{Performance Evaluation}
\label{sec:results}

\begin{figure}[h]
  \centering
  \subfigure[GARR-G Backbone]{\includegraphics[width=0.46\textwidth]{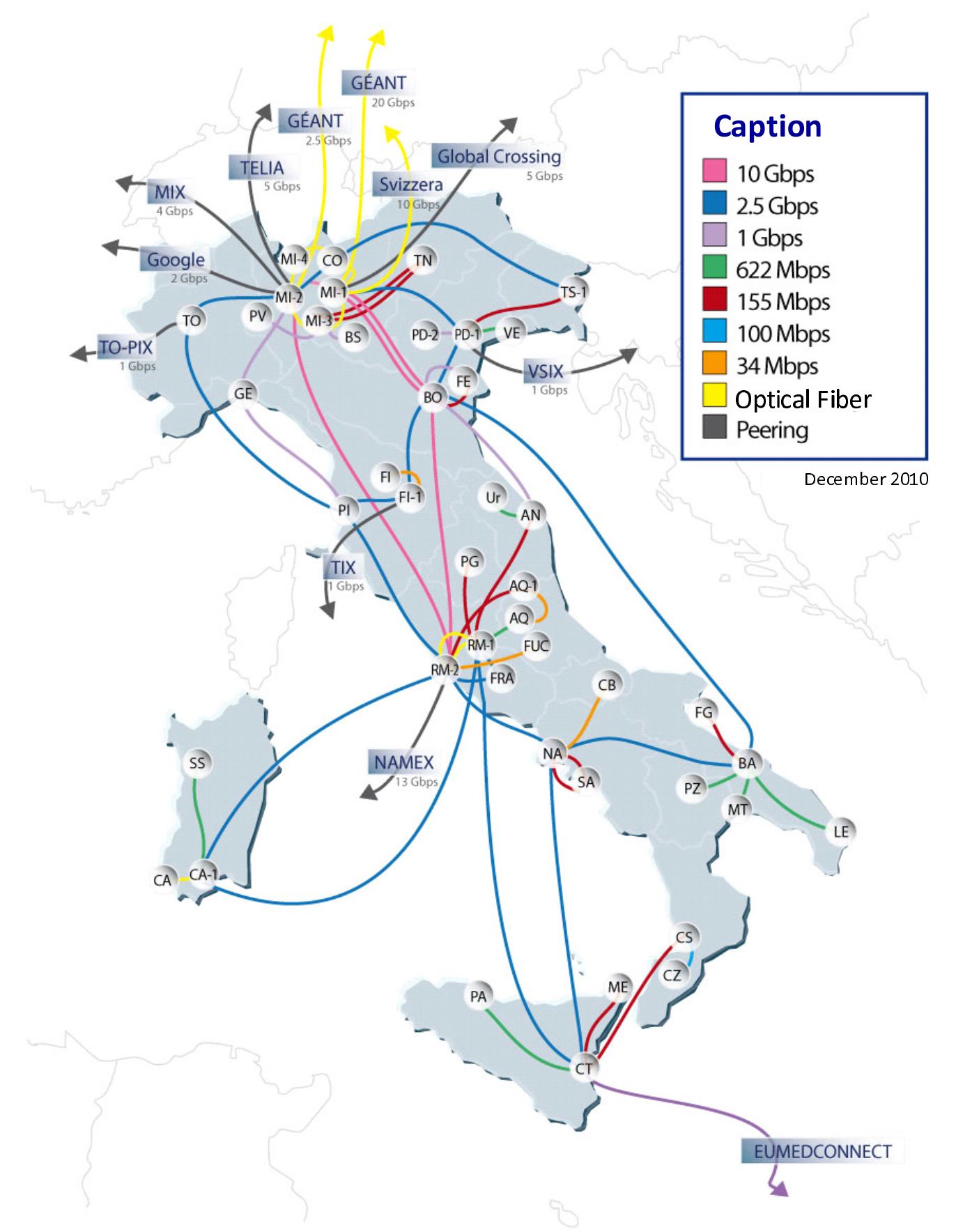}}
  \subfigure[GARR-G Topology]{\includegraphics[width=0.46\textwidth]{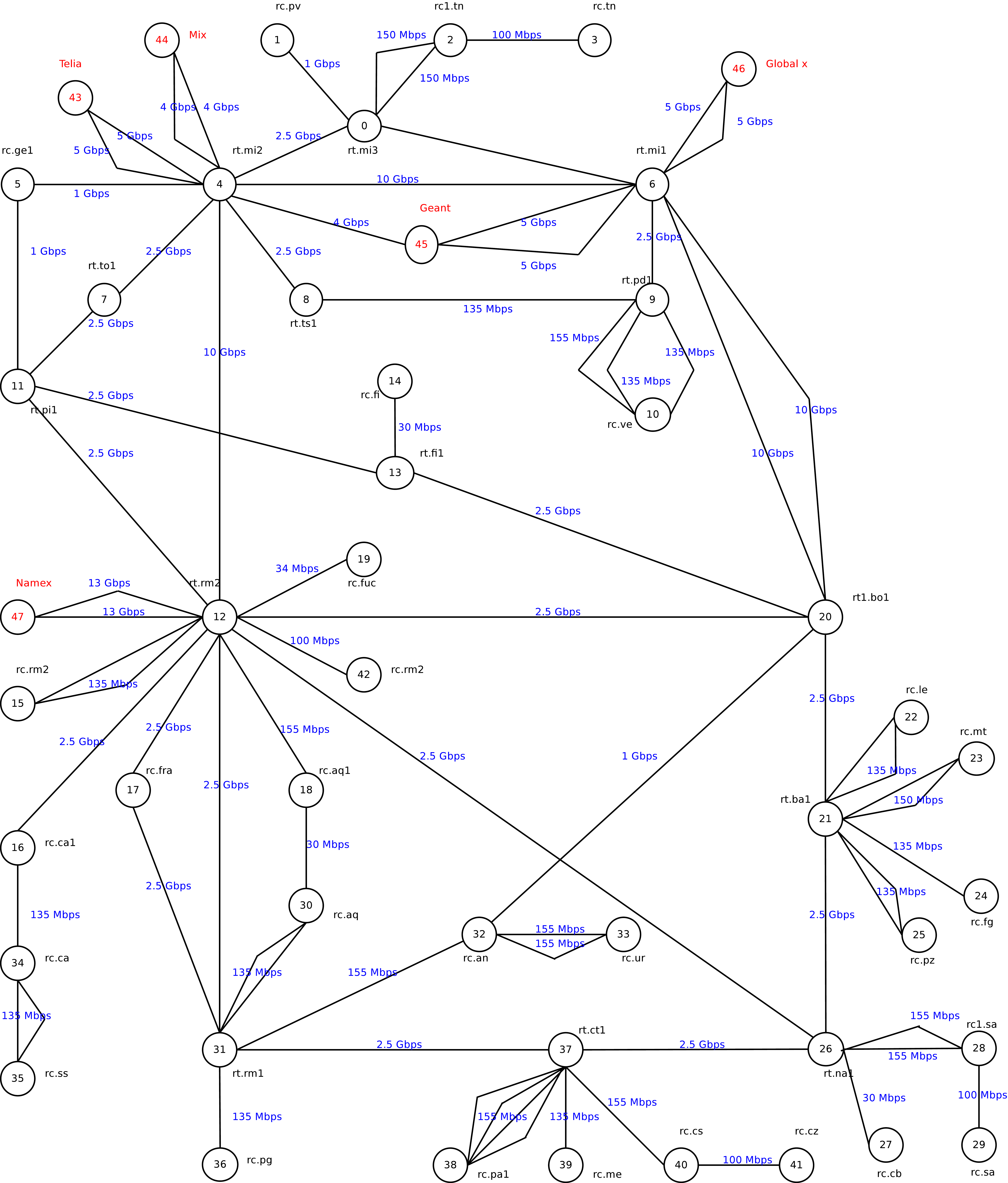}}
  \caption{The GARR research network}
	\label{fig:garr}
\end{figure}

The GOSPF protocol has been studied both in simulation and through a real-world implementation. In this section we focus on the former approach.
The algorithm proposed has been actually implemented as an extension of the OSPF routing module available in the \emph{ns-3} network simulator\footnote{\texttt{http://www.nsnam.org/}}. An exhaustive simulation campaign has been carried out in order to assess both the feasibility and the correctness of our solution, as well as to evaluate its performance in terms of energy saving and supported traffic demand.



In order to implement a realistic scenario we considered for our experiments the GARR-G backbone\footnote{\texttt{http://www.garr.it/b/eng}} (Fig.~\ref{fig:garr}(a)), the Italian Research and Education Network. From an implementation point of view, the mentioned infrastructure has been replicated in the \emph{ns3} simulator according to the topology configuration depicted in Fig.~\ref{fig:garr}(b).

\begin{figure}[h]
  \centering
  \subfigure[Daily Traffic]{\includegraphics[width=0.9\textwidth]{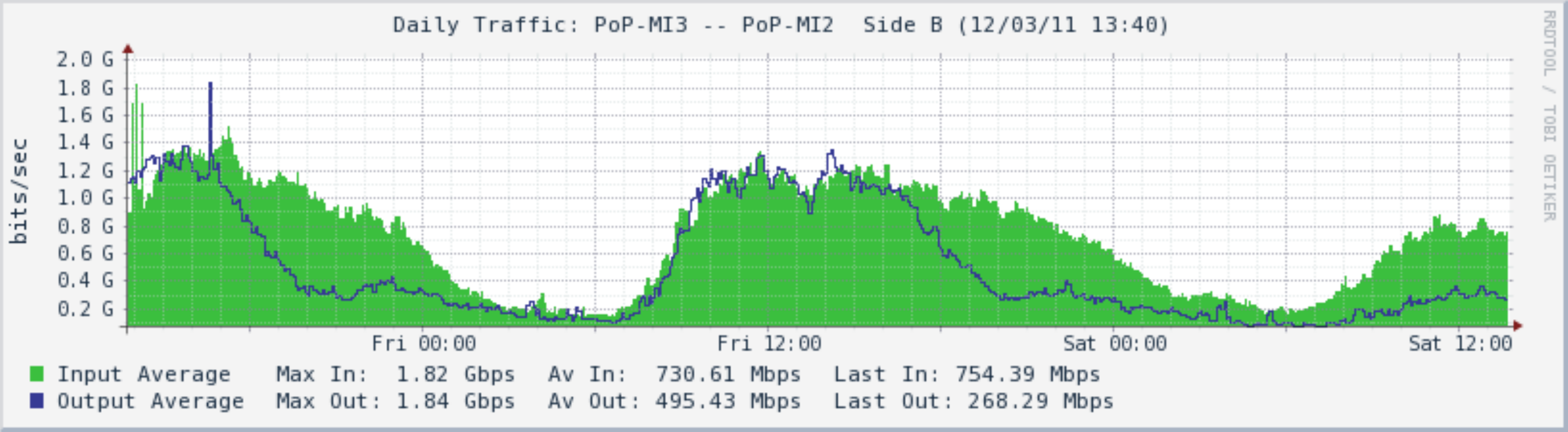}}
  \subfigure[Weekly Traffic]{\includegraphics[width=0.9\textwidth]{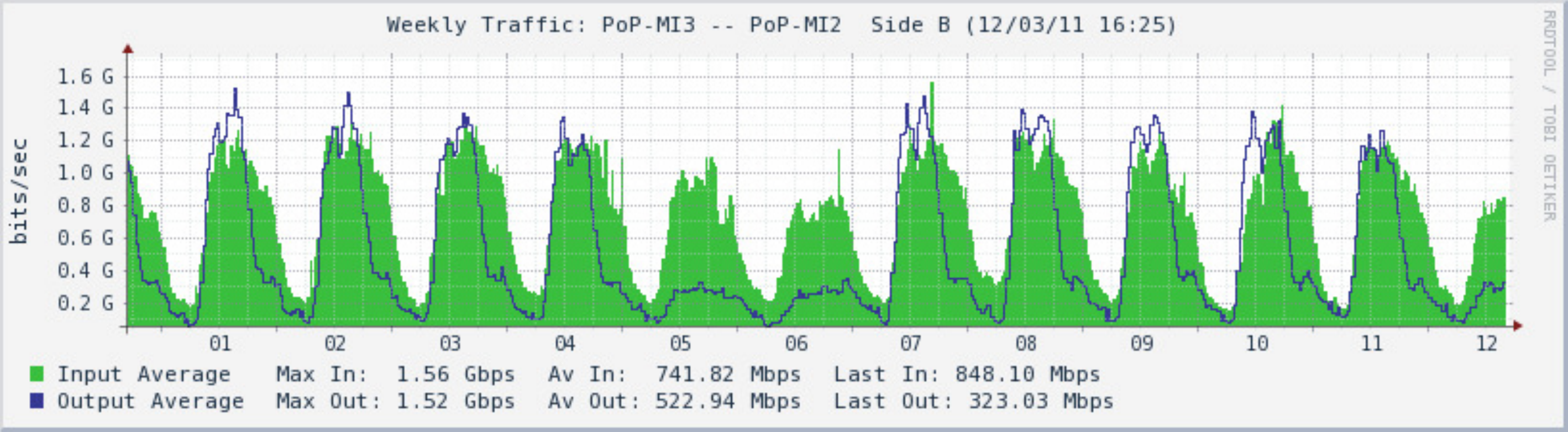}}
  \caption{Real GARR Traffic Profile}
	\label{fig:GARRTraffic}
\end{figure}

We have considered realistic traffic profiles in order to better evaluate the performance of the algorithm. In particular, we have implemented traffic patterns emulating real traffic on the GARR network. Both daily and weekly traffic profiles have been considered. A typical daily traffic profile from GARR is shown in Fig.~\ref{fig:GARRTraffic}(a). This represents the throughput, during a day, of the link between PoP Mi$3$ and PoP Mi$2$ (both located in Milan). We can see that the throughput reaches its maximum at $12$ a.m., while it decreases during night. The average link utilization is about $40$\%. Fig.~\ref{fig:GARRTraffic}(b), instead, shows the weekly throughput on the same link. We can see that the amount of traffic decreases over the weekend. All the other links in the network have a similar traffic profile.

We tested our algorithm with three different types of traffic: UDP, TCP, and mixed traffic.

\subsection{UDP Traffic}
\label{sec:UDP}

\begin{figure}[h]
  \centering
  \subfigure[Network throughput]{\includegraphics[width=0.48\textwidth]{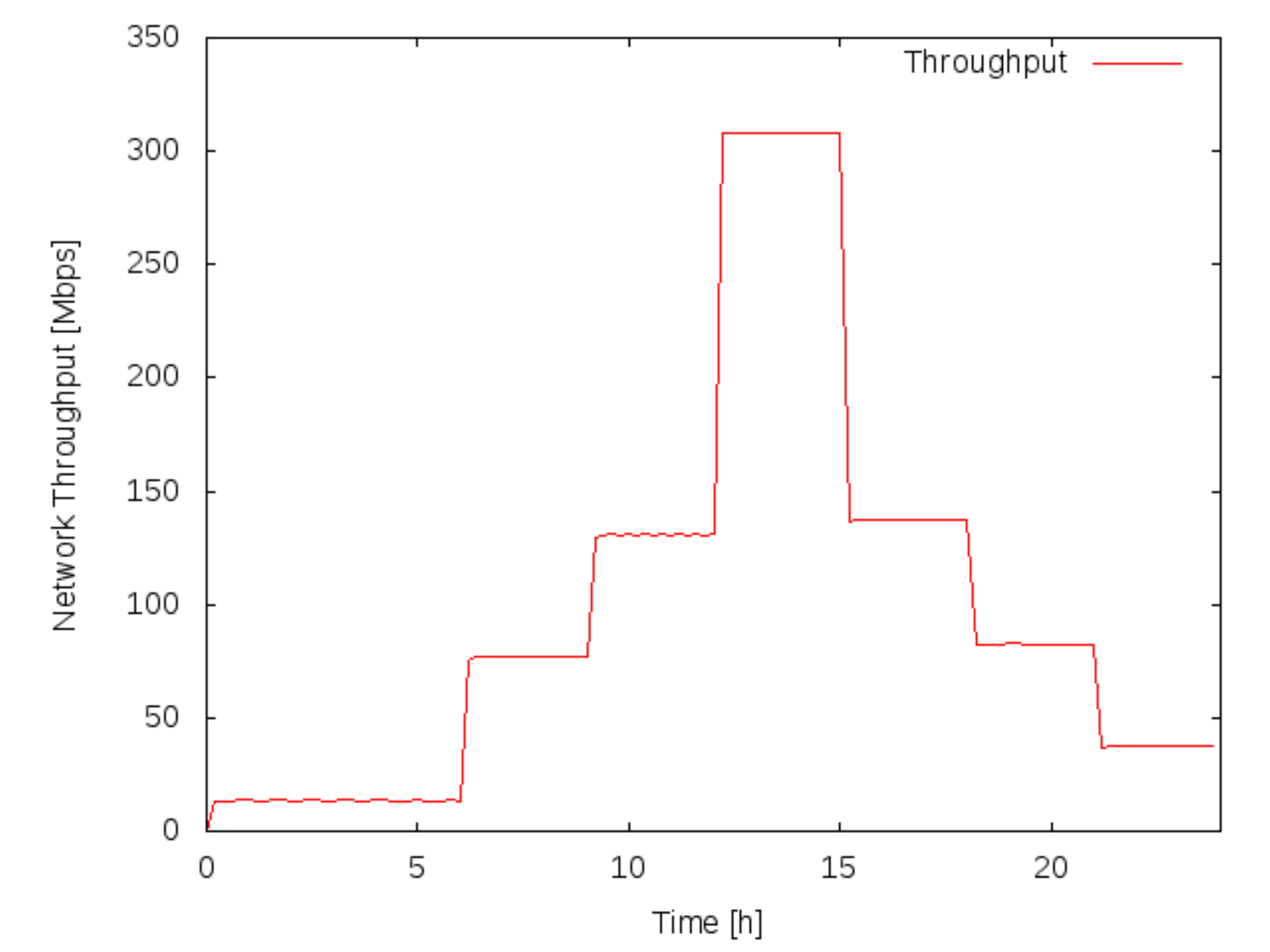}}
  \subfigure[Number of active links]{\includegraphics[width=0.48\textwidth]{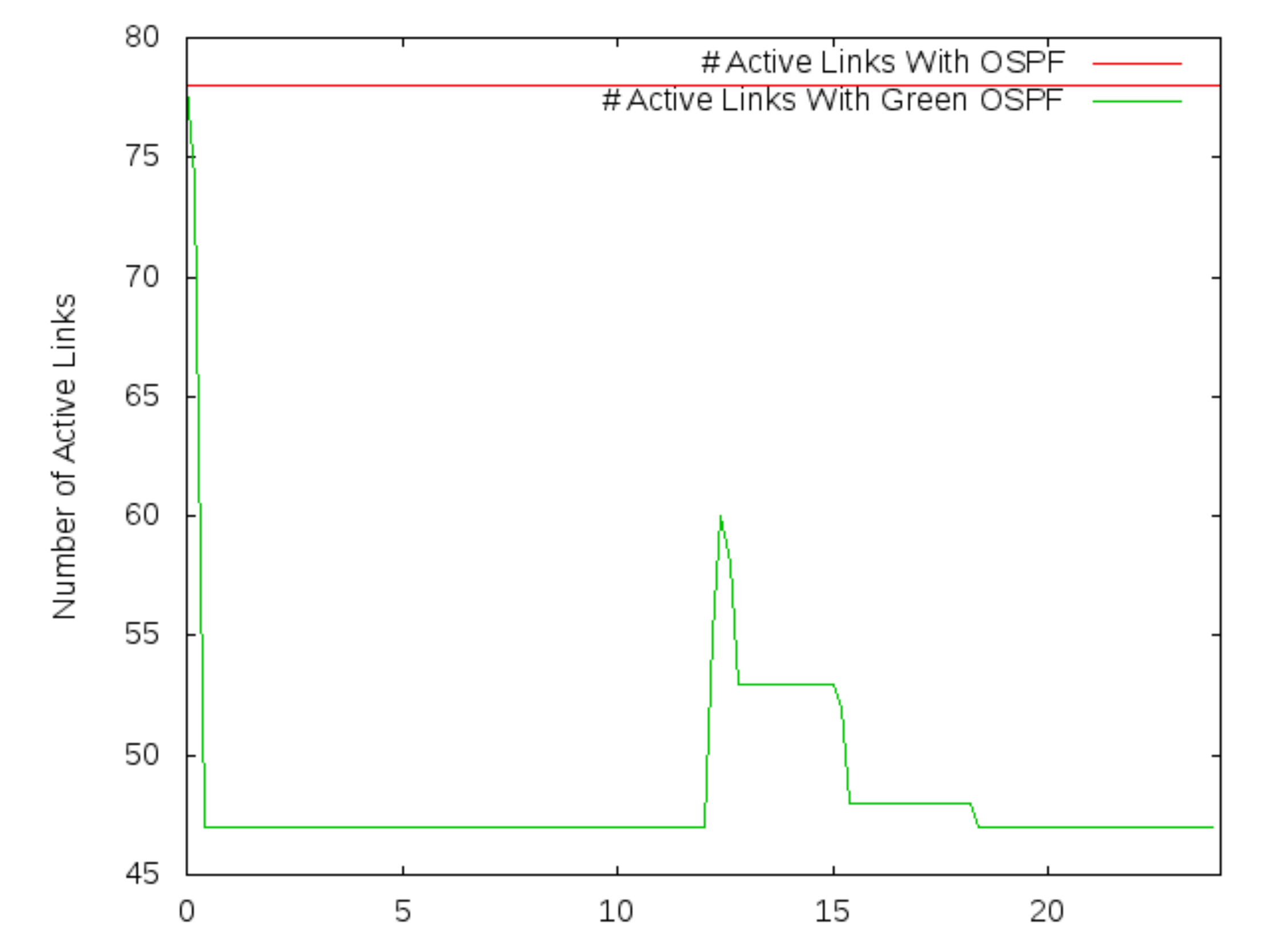}}
   \subfigure[Power consumption]{\includegraphics[width=0.48\textwidth]{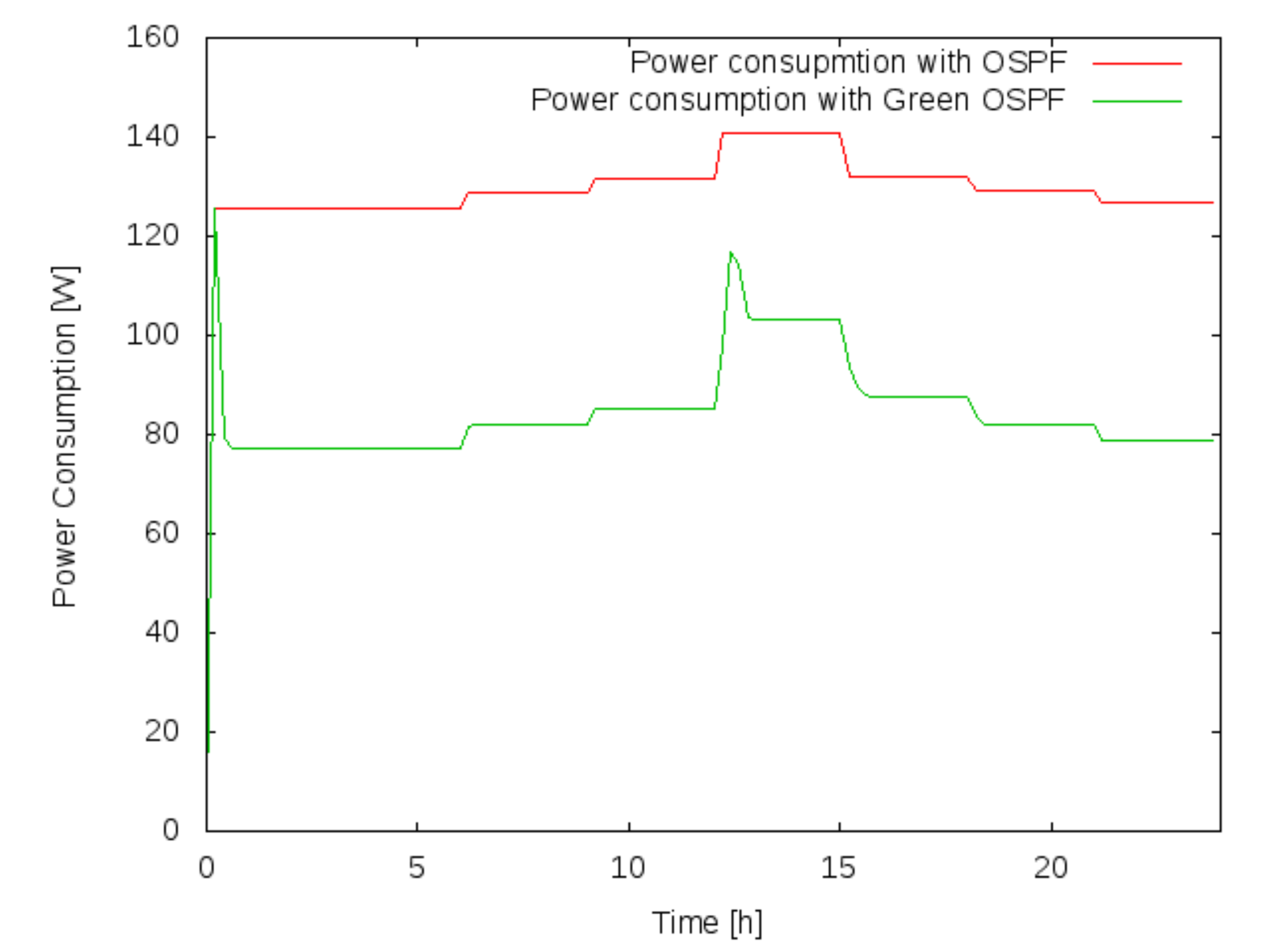}}
    \subfigure[Protocol overhead]{\includegraphics[width=0.48\textwidth]{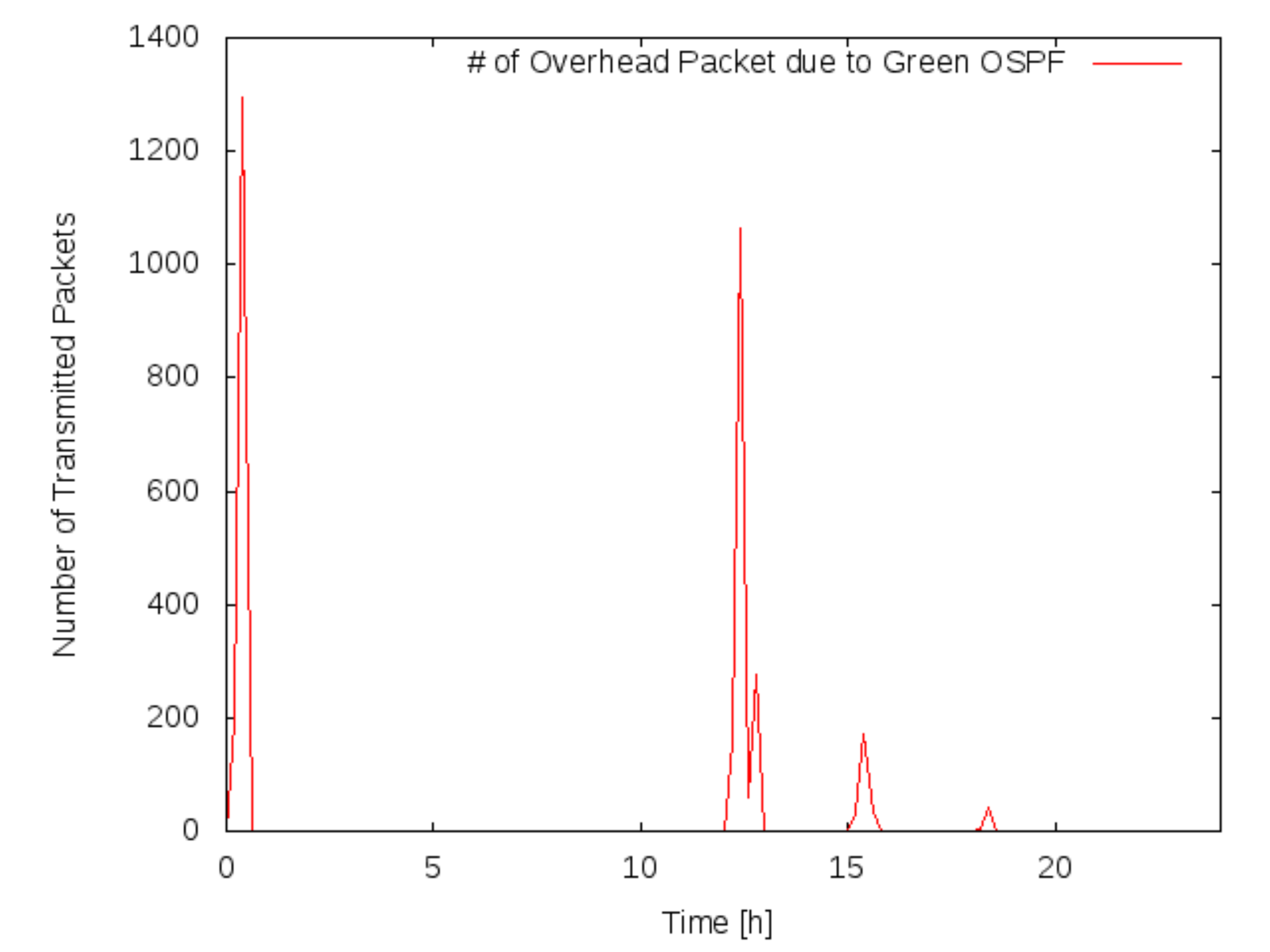}}
  \caption{Testing GoSPF with UDP}
\label{fig:GOSPF_UDP}
\end{figure}


We created a traffic profile as close as possible to the real one by considering seventeen traffic flows that are distributed in the network in such a way as to cover most links.
The flows increase their bitrate over time, until they reach the maximum between $12$ a.m. and $3$ p.m., while they decrease after midnight (see \figurename~\ref{fig:GOSPF_UDP}(a)). 
For these simulations we have set $P_{a}=1$, $P_{i}=0.8 $, $P_{s}=0.016$, $\gamma _{u}=20$\%, and $\gamma _{l}=80$\%.

With respect to the above mentioned point, the energy model proposed in equation~(\ref{eq:energy}) has been implemented in ns$3$. This required us to modify three classes: (i) \texttt{PointToPointNetDevice}, which models a generic point-to-point device or serial link; (ii) \texttt{Node}, representing a node in the network; (iii) \texttt{Node Container}, which, as the name itself suggests, contains all nodes in the network. Energy consumption in the network is computed by the sum of the energy consumptions of all interfaces. To compute this last item, a new member function, called \texttt{GetDEnergy} , has been added to the \texttt{PointToPointNetDevice} class. The \textbf{Node} class has in turn been modified with the insertion of a \texttt{Node::GetEnergy} method which allows to compute the sum of the energy consumptions of the \texttt{PointToPointNetDevice} objects associated with it. In the same way, the overall energy consumption is obtained via the newly created \texttt{NodeContainer::GetTotalEnergy} method of the \texttt{NodeContainer} class, which basically provides the sum of the previously mentioned values for each node.

\subsubsection{Number of Active UDP Links}
\label{subsubsec:num active links U}


\begin{figure}
  \centering
  	\includegraphics[width=0.65\columnwidth]{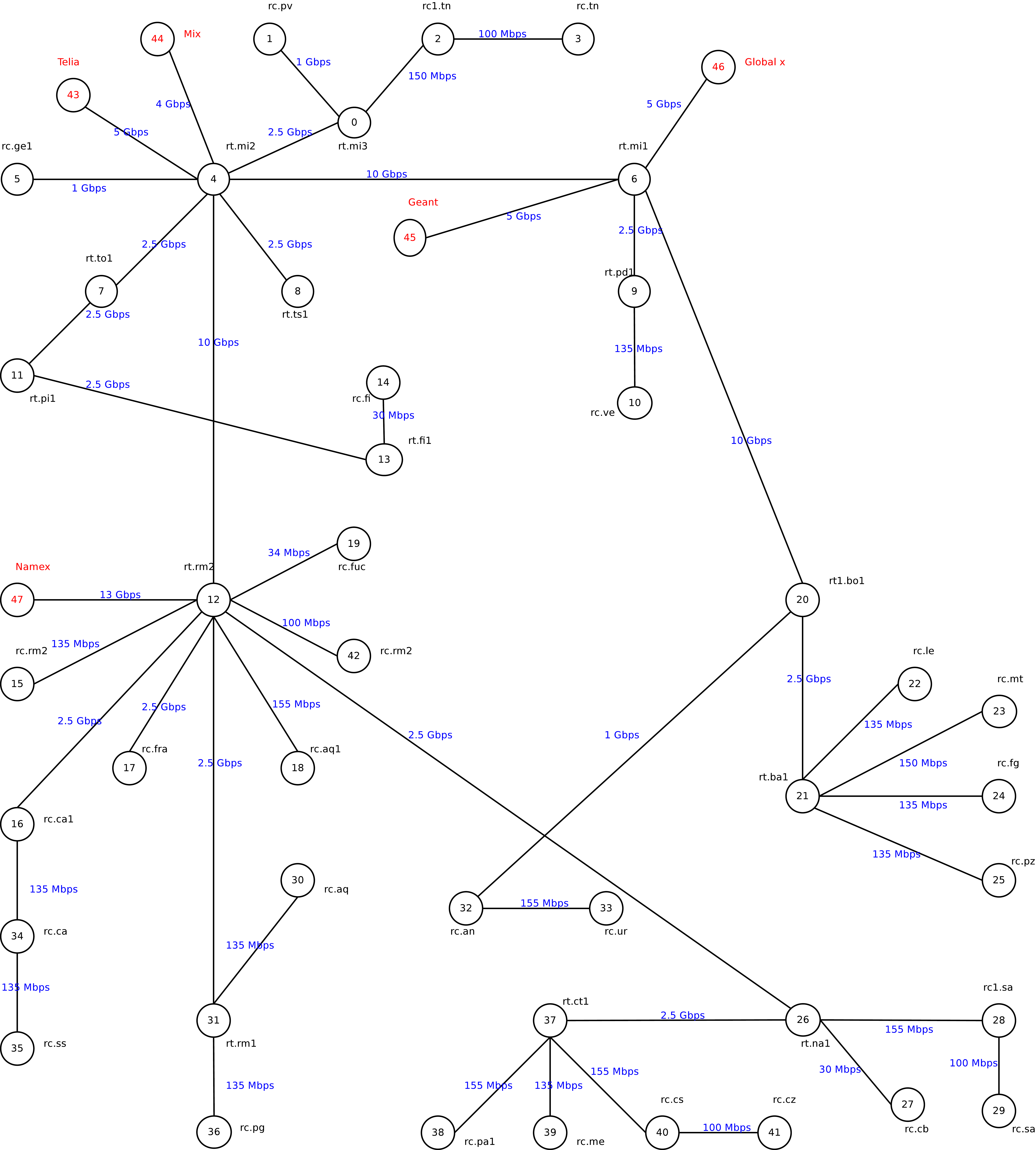}\\
  \caption{Minimum GARR-G Topology}\label{fig:garrTopoMin}
\end{figure}

Fig.~\ref{fig:GOSPF_UDP}(b) shows the number of active links over time. We can see that until $12$ a.m. the number of active links is constant and equal to $47$. Such a value represents the minimum number of links to ensure connectivity, and is in fact equal to $ |V|-1 $. In this phase the network topology is represented by the MCST shown in Fig.~\ref{fig:garrTopoMin}.  At $12$ a.m. the number of active links has a peak. In fact the overall throughput has become too high and the MCST is not able to support it anymore. The peak triggers the graft process which restores all links located at one hop distance from overloaded links. The subsequent checks on the utilization rate show that some links are actually used while others are not. Between $12$ a.m. and $3$ p.m. the number of active links is $54$; such number then becomes $48$ and, eventually, falls back again to $47$. We can appreciate an abrupt initial decrease in the number of active links. Indeed, in the beginning all links are active and the overall network throughput is very low; consequently, when the algorithm starts, it cuts all links that do not belong to the MCST.

\subsubsection{UDP Energy Consumption}
\label{subsubsec:En Co}


The results show, with the traffic profile depicted in Fig.~\ref{fig:GOSPF_UDP}(a), a total energy saving of about $34,8$\%. Fig.~\ref{fig:GOSPF_UDP}(c) shows the network power consumption over time, which can be justified through a comparison with Fig.~\ref{fig:GOSPF_UDP}(b) (showing the number of active links).
Initially, power consumption is similar to the case in which our algorithm is not employed, since all links are active.
As we can see from the comparison between these two figures and Fig.~\ref{fig:GOSPF_UDP}(a), energy consumption increases its value as long as both the traffic and the number of active links increase.
In fact, the percentage of energy saved decreases until it reaches its minimum between $12$ a.m. and $3$ p.m.; after this interval it increases again.
Note that both the number of active links and the energy consumption depend on the specific features of the network traffic transported across the network.

\subsubsection{Other Parameters}
\label{subsubsec:O P}

We also consider \emph{packet losses}, as well as the \emph{overhead} introduced by our algorithm. For the UDP daily traffic case, packet losses do not increase with respect to the value obtained by simulation in the absence of our algorithm. With respect to the overhead induced by our algorithm, we observe that graft and cut update packets represent about $0,8$\% of the overall network traffic. Initially we can note the highest peak due to the high number of cuts. Flooding of these packets causes also an increase in power consumption; in fact from the comparison between Fig.~\ref{fig:GOSPF_UDP}(c) and Fig.~\ref{fig:GOSPF_UDP}(d), we can note that when overhead packets reach their peak value, due to grafting and cutting operations, the energy saved decreases accordingly.


\begin{table}[h!]
\caption{Daily UDP Traffic Test}
\centering
\begin{tabular}{l l l}
\hline
-- & Green OSPF &  OSPF \\
\hline
Total Energy Consumption & 2035.94 J & 3121.38 J \\
\hline
Average Number of Active Links & 48.517 & 78   \\
\hline
Packet Loss & 0\% & 0\% \\
\hline
Overhead\% & 0.810\% & 0\% \\
 \hline
\end{tabular}
\end{table}

\subsection{TCP Traffic}
\label{sec:TCP d}

The traffic profile adopted with TCP is similar to the one we considered in the UDP case. Also the values of the variables $P_{a}$, $P_{i}$, $P_{s}$, $\gamma _{u}$ and $\gamma _{l}$ are the same as in the UDP traffic test.

\begin{figure}[h]
  \centering
  \subfigure[Total network throughput]{\includegraphics[width=0.48\textwidth]{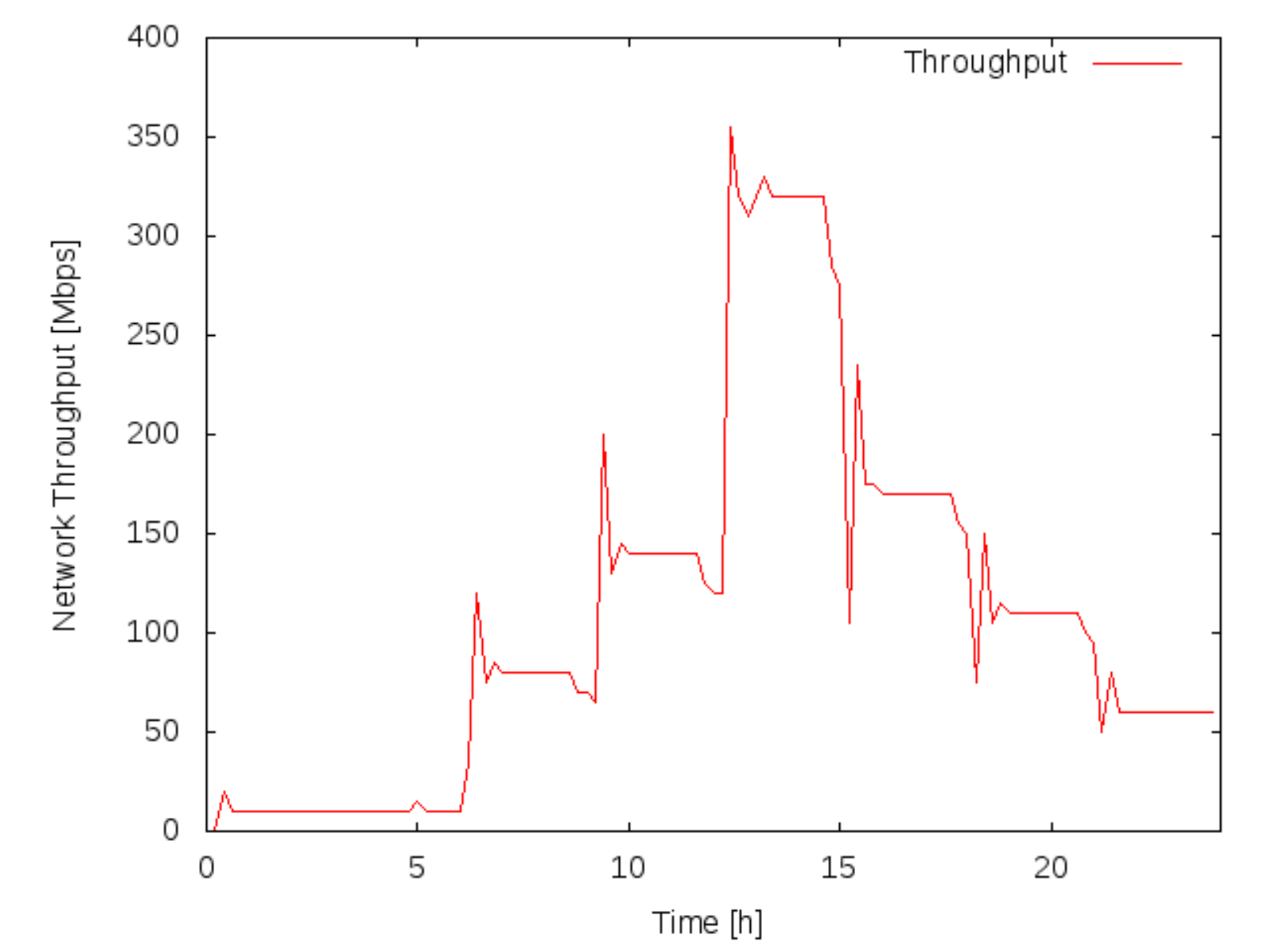}}
  \subfigure[Number of active links]{\includegraphics[width=0.48\textwidth]{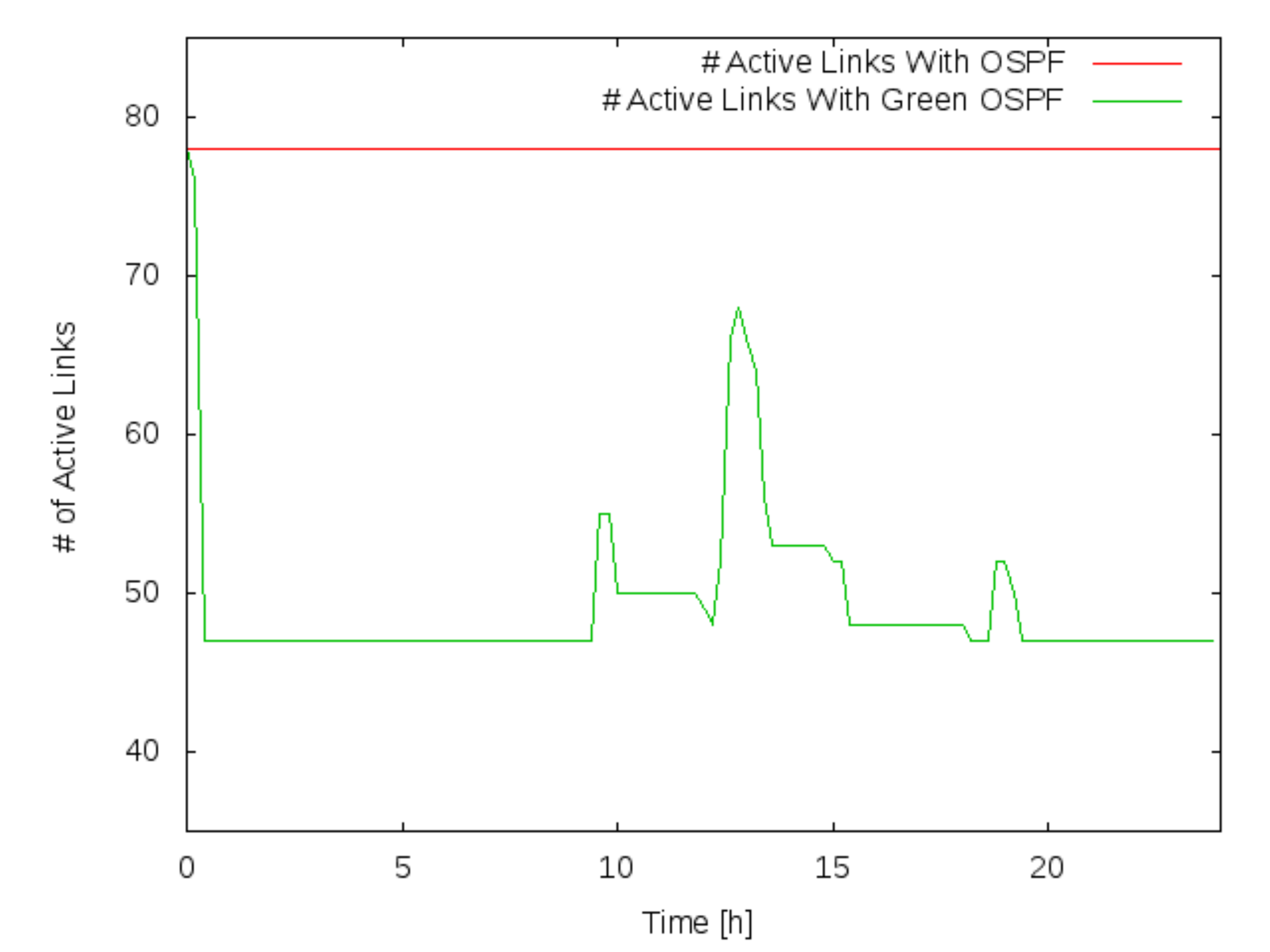}}
   \subfigure[Power consumption]{\includegraphics[width=0.48\textwidth]{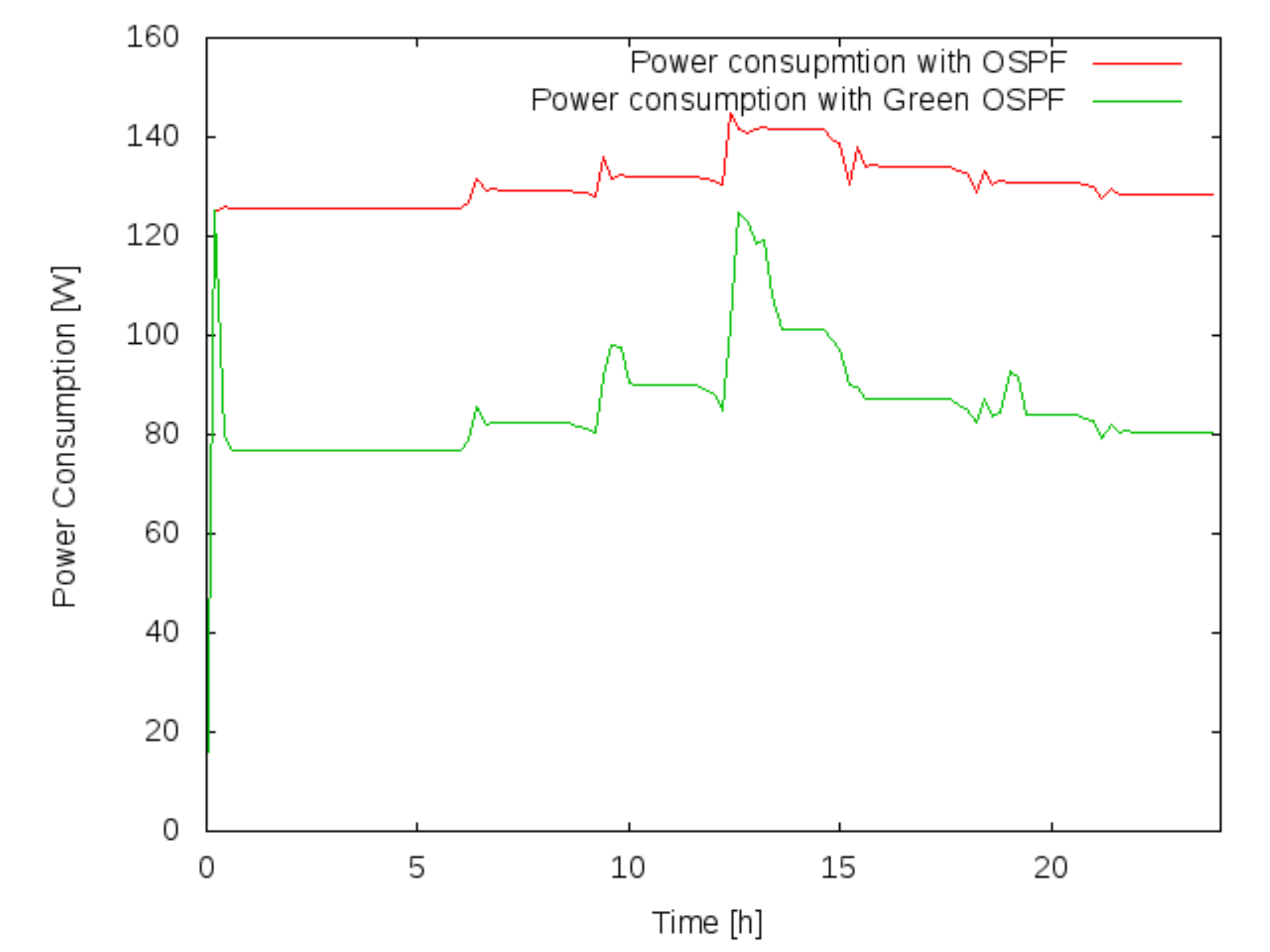}}
  \caption{Testing GoSPF with TCP}
\label{fig:GOSPF_TCP}
\end{figure}





We can see in Fig.~\ref{fig:GOSPF_TCP}(a) that there are some traffic peaks due to the establishment of new TCP connections. The graph in Fig.~\ref{fig:GOSPF_TCP}(b) shows the daily number of active links in the network. We can note the difference between the UDP and TCP traffic cases. Initially, the number of active links grows and shows three peaks corresponding to those that occur in Fig~\ref{fig:GOSPF_TCP}(a). In fact, the spikes in the network throughput, due to new TCP connections, produce sudden increases in the utilization rate of some links that in turn trigger grafting operations. We also remark that the links which are powered on between $9$ a.m. and $12$ a.m. are in this case $50$ (against $47$ as in the UDP case) because the grafting operation restores some links whose utilization level is above the threshold value of $20$\%.


Coming to energy consumption, the results show an overall energy saving of about $34,2$\%. With respect to the UDP case, we can note a slight decrease of this value, due to the higher number of active links, as well as supported traffic. We finally remark how packet losses increase in the TCP case, moving from $0.020$\% to $0.026$\%. The same holds true for overhead packets, whose percentage increases (with respect to the UDP case) from $0,80$\% to $0,84$\% of the useful traffic.

\begin{table}[h!]
\caption{Daily TCP Traffic Test}
\centering
\begin{tabular}{l l l}
\hline
-- & Green OSPF & OSPF \\
\hline
Total Energy Consumption & 2063.86 J & 3138.44 J \\
\hline
Average Number of Active Links & 49.250 & 78   \\
\hline
Packet Loss & 0.026\% & 0.020\% \\
\hline
Overhead\% & 0.844\% & 0\% \\
 \hline
\end{tabular}
\end{table}

\subsection{Mixed TCP and UDP Traffic}
\label{sec:TCP UDP d}

\begin{figure}[h]
  \centering
  \subfigure[Power consumption]{\includegraphics[width=0.48\textwidth]{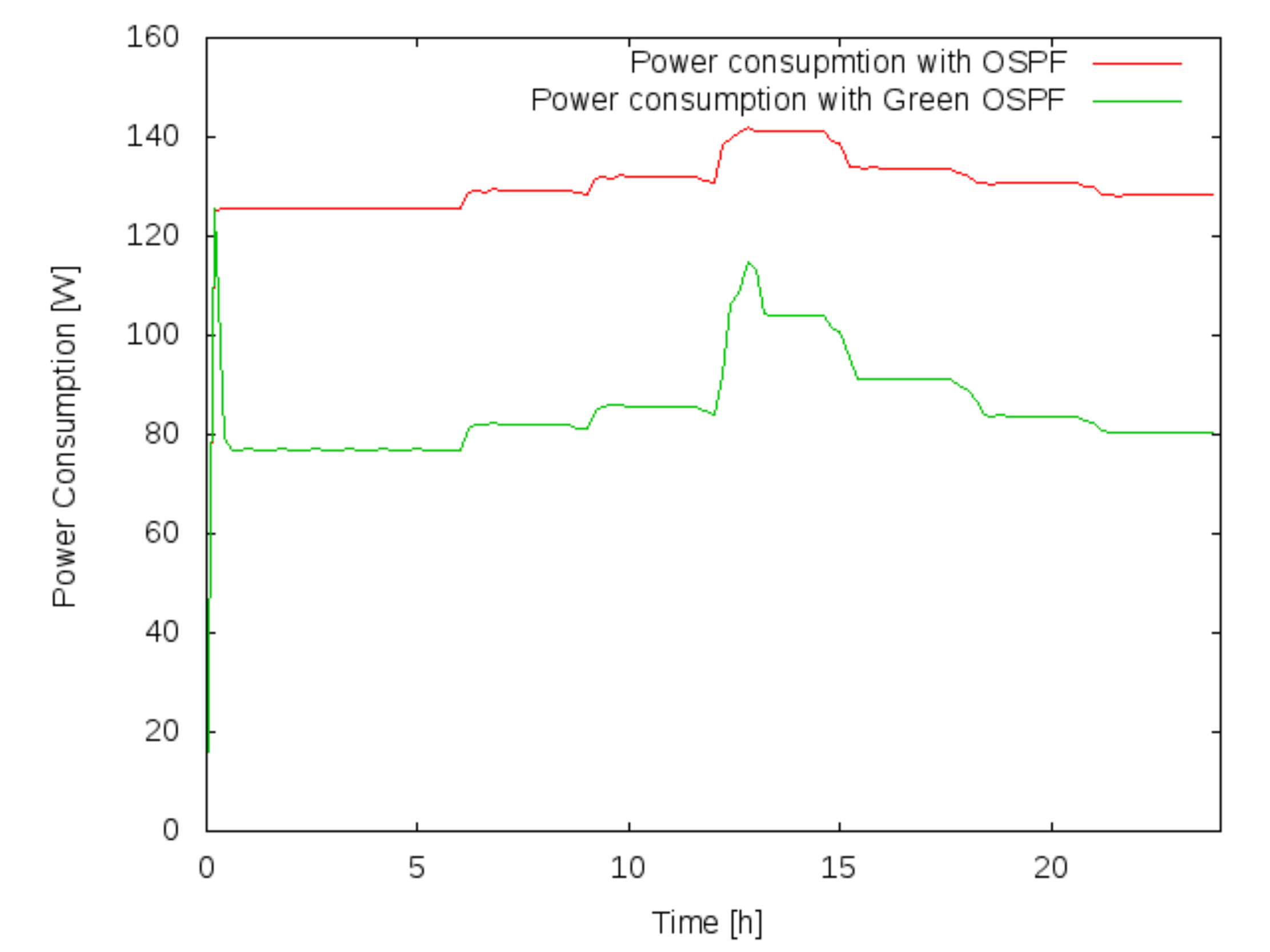}}
  \subfigure[\% of overhead packets]{\includegraphics[width=0.48\textwidth]{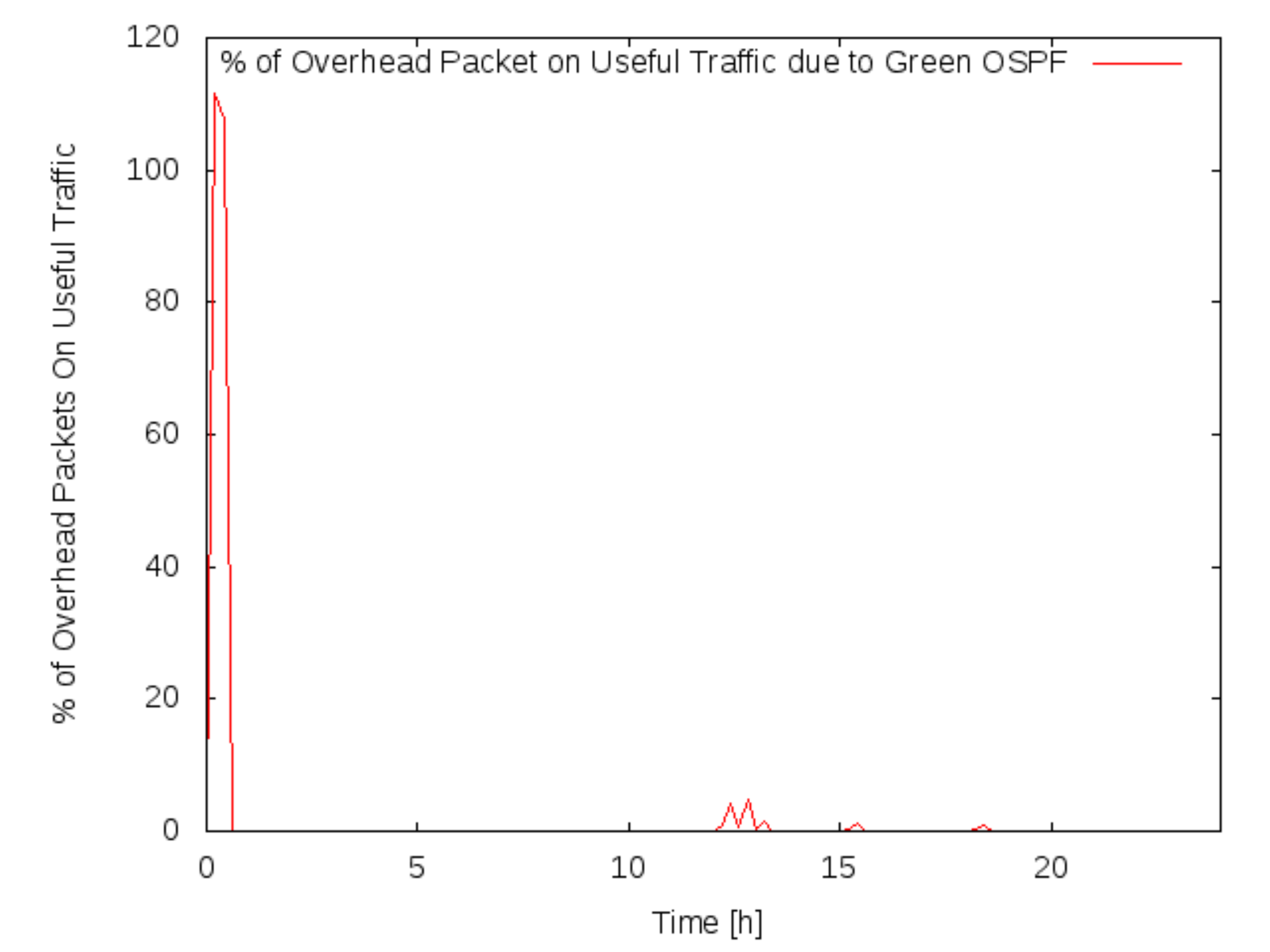}}
  \caption{Testing GoSPF with a mixed traffic profile}
\label{fig:GOSPF_Mixed}
\end{figure}

We finally consider a daily traffic composed of both TCP and UDP flows. The values of the variables $P_{a}$, $P_{i}$ ,$P_{s}$, $ \gamma _{u}$, $ \gamma _{l}$ and $T_{sample}$ are the same as those we adopted for the other tests. 
The number of active links during time is very similar to the UDP traffic test. The energy consumed (see Fig.~\ref{fig:GOSPF_Mixed}(a)) is $34,5$\% of the overall network energy consumption in the absence of our algorithm. There is an increase in packet losses between the case when our algorithm is activated and the case when it is switched off; losses go from $0.010$\% to $0.015$\%. The percentage of overhead packets over useful traffic (Fig.~\ref{fig:GOSPF_Mixed}(b)) is about $0.44$\%, which is lower than that observed in the TCP test.

%

\begin{table}[h!]
\caption{Daily UDP and TCP Traffic Test}
\centering
\begin{tabular}{l l l}
\hline
-- & Green OSPF & OSPF \\
\hline
Total Energy Consumption & 2054.05 J & 3135.7 J \\
\hline
Average Number of Active Links & 48.650 & 78   \\
\hline
Packet Loss & 0.015\% & 0.010\% \\
\hline
Overhead\% & 0.442\% & 0\% \\
 \hline
\end{tabular}
\end{table}

Table~\ref{tab:summary} summarizes the above discussed results and helps the reader make a comparison between the three scenarios we have so far analyzed.

From an analysis of the data in the table, we can draw some interesting considerations. First, it looks clear that GOSPF outperforms OSPF in all of the considered traffic scenarios. Indeed, at the price of an almost negligible overhead, GOSPF allows us to obtain substantial gains in terms of energy efficiency. Furthermore, it is worth to remark the impact that the transport-layer protocol can have on the overall `energy performance' of the network. The differences between the TCP and UDP traffic cases can in fact be appreciated in all of the trials we conducted. Such a difference can be ascribed to the transport-layer overhead induced by TCP, which is particularly relevant at connection setup time (i.e., when TCP performs the initial three-way handshake).  TCP signaling overhead happens to produce non-negligible increases in the utilization rate of some links that in turn trigger grafting operations, with the final result of keeping active a higher number of links than those needed in the UDP case.

\begin{table}[h!]
\caption{Comparison Between Daily UDP, TCP ,UDP and TCP Traffic Test}
\label{tab:summary}
\centering
\begin{tabular}{p{1.5cm} p{1.5cm} p{1.5cm} p{1.5cm}}
\hline
-- & UDP Traffic & TCP Traffic & UDP and TCP Traffic \\
\hline
Total Energy Saving & 34,8\% & 34,2\% & 34,5\% \\
\hline
Average Number of Active Links & 48.517 & 49.250 & 48.650  \\
\hline
Packet Loss & 0\% & 0.026\% & 0.015 \% \\
\hline
Overhead\% & 0.810\% & 0.844\% & 0.442\% \\
 \hline
\end{tabular}
\end{table}

\subsection{UDP Traffic Test with a higher sampling frequency}
\label{sec:UDP F}

\begin{figure}[h]
  \centering
  \subfigure[Number of active links]{\includegraphics[width=0.48\textwidth]{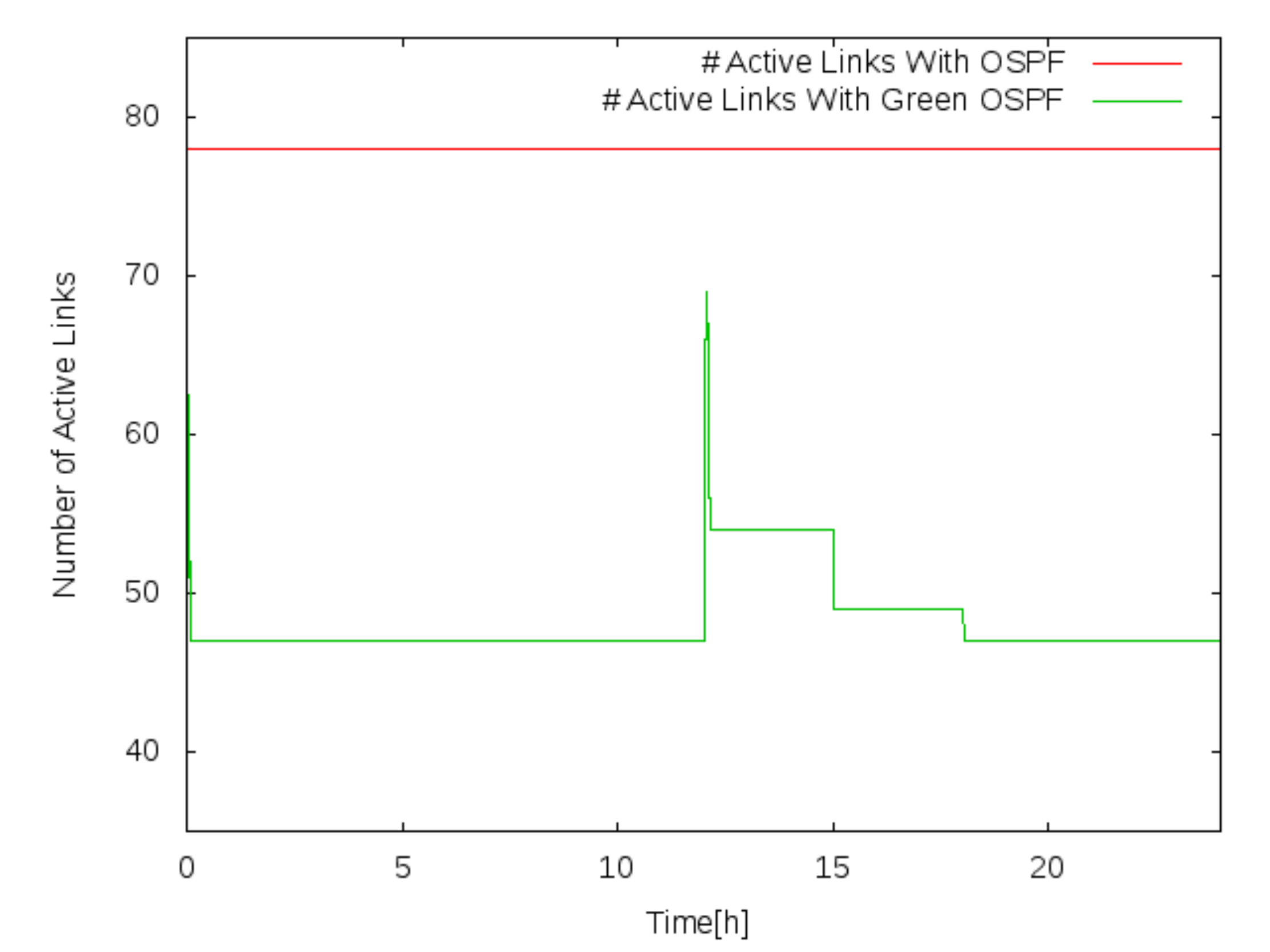}}
  \subfigure[Power consumption]{\includegraphics[width=0.48\textwidth]{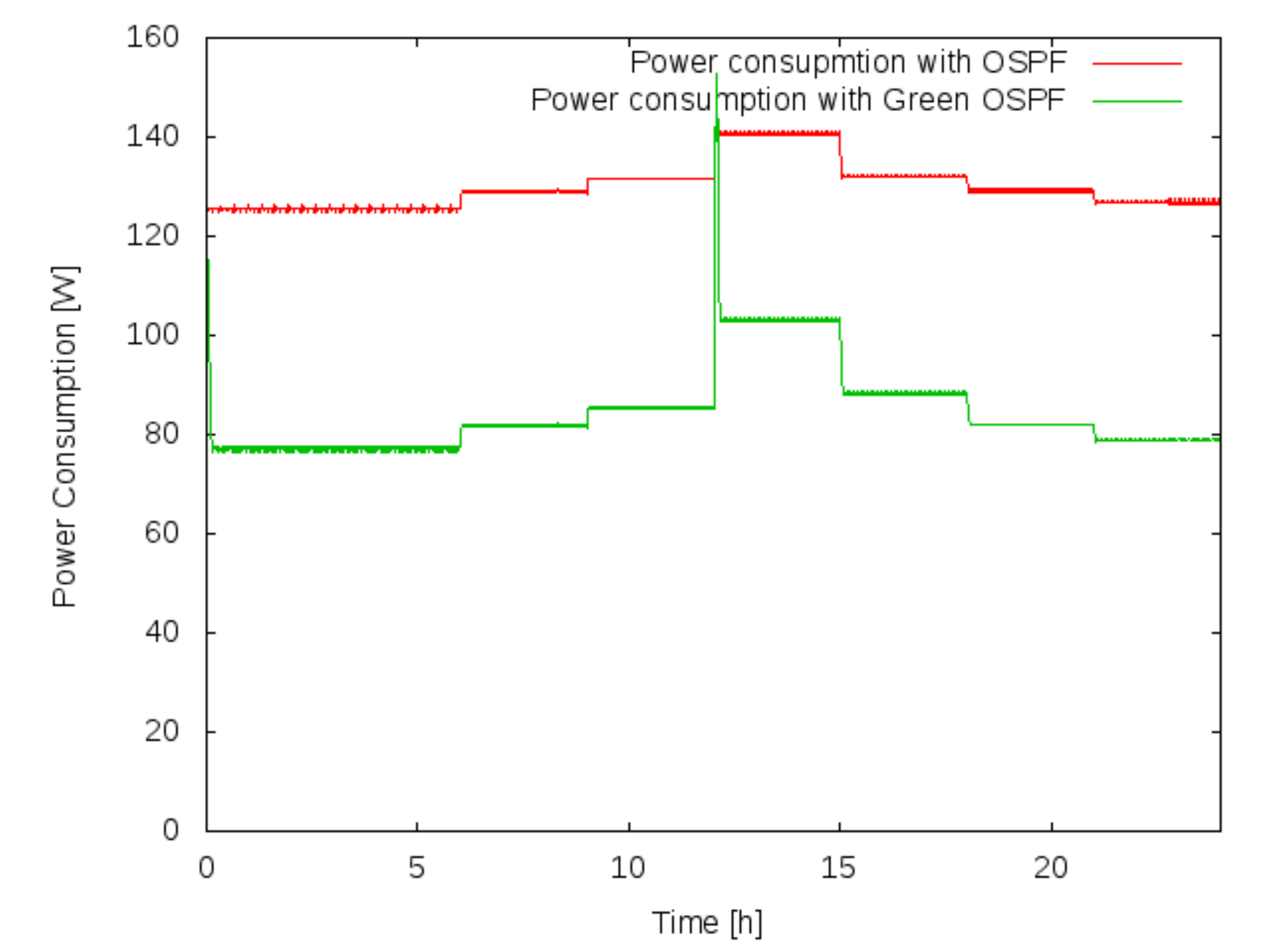}}
  \caption{UDP test with a smaller sampling period}
\label{fig:GOSPF_Frequency}
\end{figure}

The sampling period is an important parameter which the performance of our algorithm depends upon.
In the following we show the results that we obtain in the case of a higher sampling frequency, namely $T_{sample}=0.02s$,  considering both a daily UDP traffic and a daily mixed traffic.
The first graph (Fig.\ref{fig:GOSPF_Frequency}(a)) shows the number of active links, while Fig.\ref{fig:GOSPF_Frequency}(b) shows the network power consumption.
These are similar to Fig.~\ref{fig:GOSPF_UDP}(c) and Fig.~\ref{fig:GOSPF_UDP}(b), representing, respectively, the power consumption  and the number of links that are powered on with $T_{sample}=0.2s$. From Table~\ref{tab:sampling_1} we observe, for the mixed traffic case, a clear increase in the number of overhead packets, reaching $2$\% of the useful traffic. In this scenario there is also an increase in packet losses, which pass from $0.015$\% to $0.096$\%. The increase in the number of overhead packets is caused by a greater number of check operations at the nodes and hence a greater number of cuts or grafts. We can imagine that with a smaller $T_{sample}$ a node detects sooner the overload of an interface; though, the higher number of overhead packets, with the unavoidable increase of throughput in the network, has the side-effect of increasing packet losses.

Energy saving with $T_{sample}=0.02s$ is about $35,06$\%, which is slightly higher than the value obtained with $T_{sample}=0.2s$  (see Table~\ref{tab:sampling_2}). We have this result because a node realizes faster that an interface is underutilized. Moreover, from the comparison between Fig.~\ref{fig:GOSPF_Frequency}(a) and Fig.~\ref{fig:GOSPF_UDP}(b) we can see a higher peak at $12$ a.m.: since $T_{sample}$ represents the time interval over which the \textit{activity time} $T_{ac}$ is averaged, a smaller sample time causes a lower immunity to sudden spikes. In particular at $12$ a.m. there is a flooding of Graft and Cut Update Packets that leads to a peak of the utilization rate $U_{rate}$ on some links.

%

\begin{table}[h]
\caption{Daily UDP and TCP traffic tests with two different sampling periods}
\label{tab:sampling_1}
\centering
\begin{tabular}{l l l}
\hline
-- & $T_{sample}=0,2s$ &  $T_{sample}=0,02s$\\
\hline
Packet Loss & 0.015 \% & 0.096 \%  \\
\hline
Overhead\% & 0.442\% & 2,012 \%  \\
 \hline
\end{tabular}
\end{table}

\begin{table}[h!]
\caption{Daily UDP traffic tests with two different sampling periods}
\label{tab:sampling_2}
\centering
\begin{tabular}{l l l}
\hline
-- & $T_{sample}=0,2s$ & $T_{sample}=0,02s$\\
\hline
Total Energy Saving & 34,8\% & 35,06\%  \\
\hline
Average Number of Active Links & 48.517 & 48.236   \\
\hline
\end{tabular}
\end{table}

\subsection{UDP Weekly Traffic Test}
\label{sec:UDP W}

\begin{figure}[h]
  \centering
  \subfigure[Network throughput]{\includegraphics[width=0.48\textwidth]{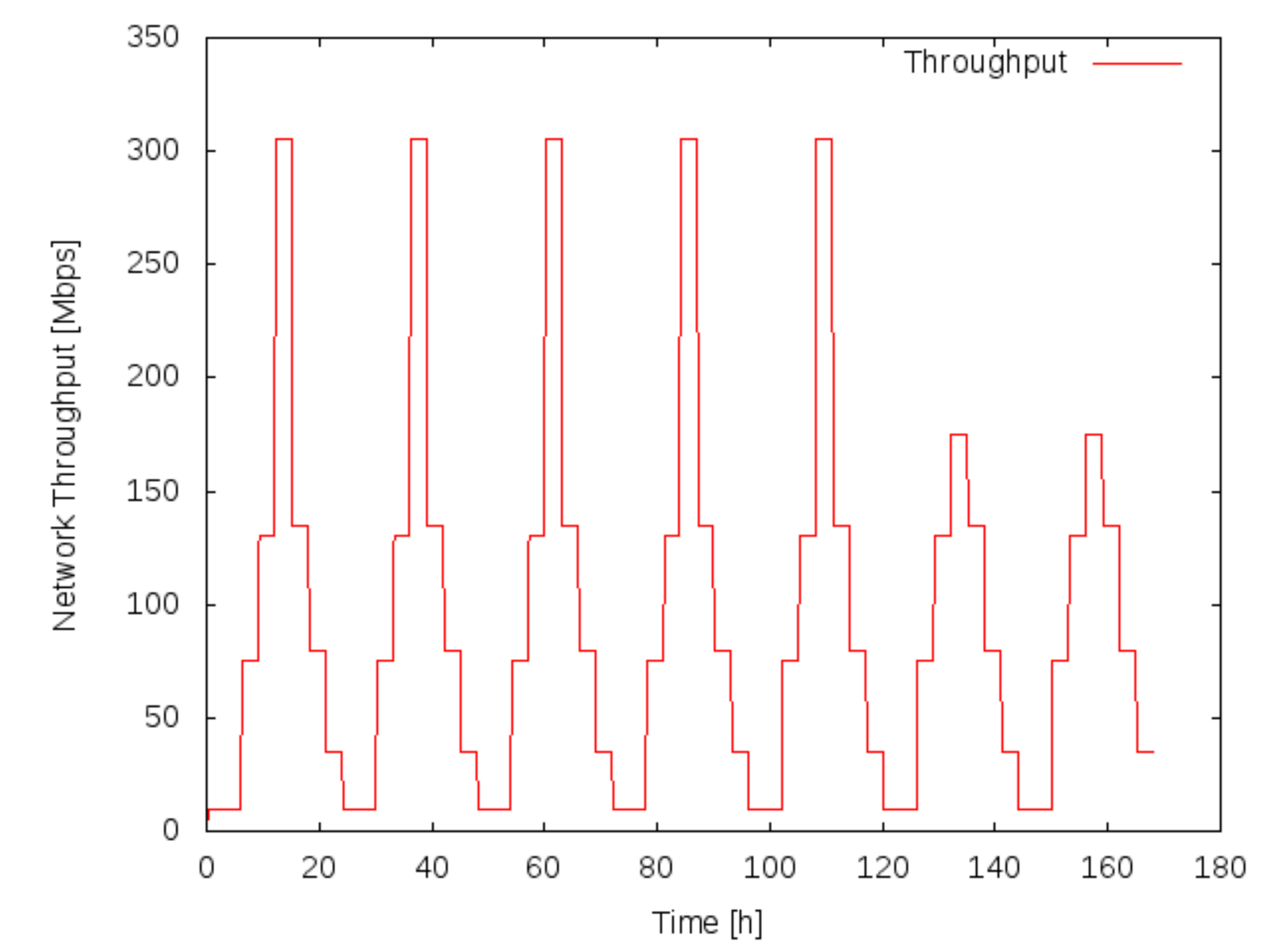}}
  \subfigure[Active links]{\includegraphics[width=0.48\textwidth]{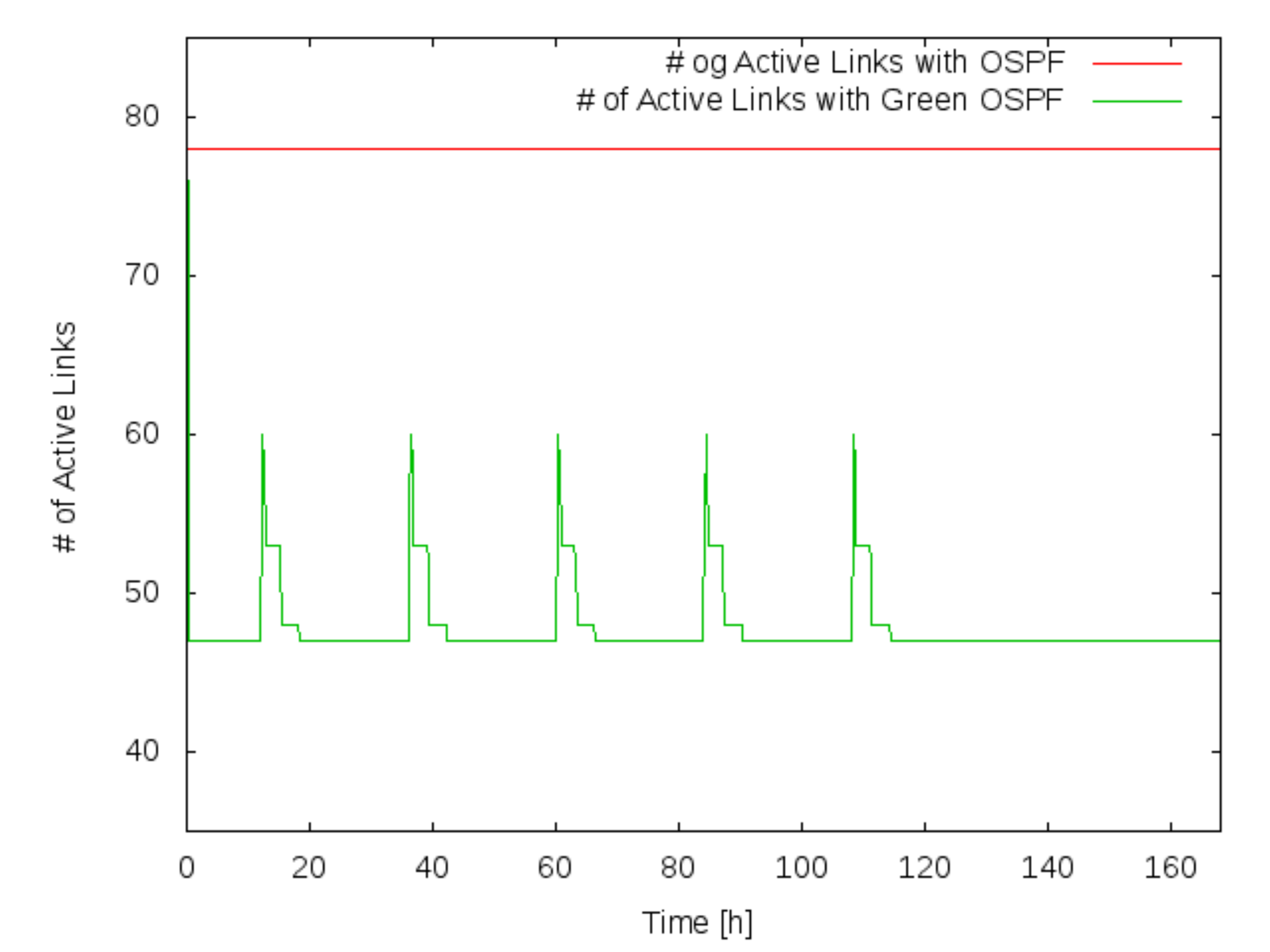}}
    \subfigure[Power consumption]{\includegraphics[width=0.48\textwidth]{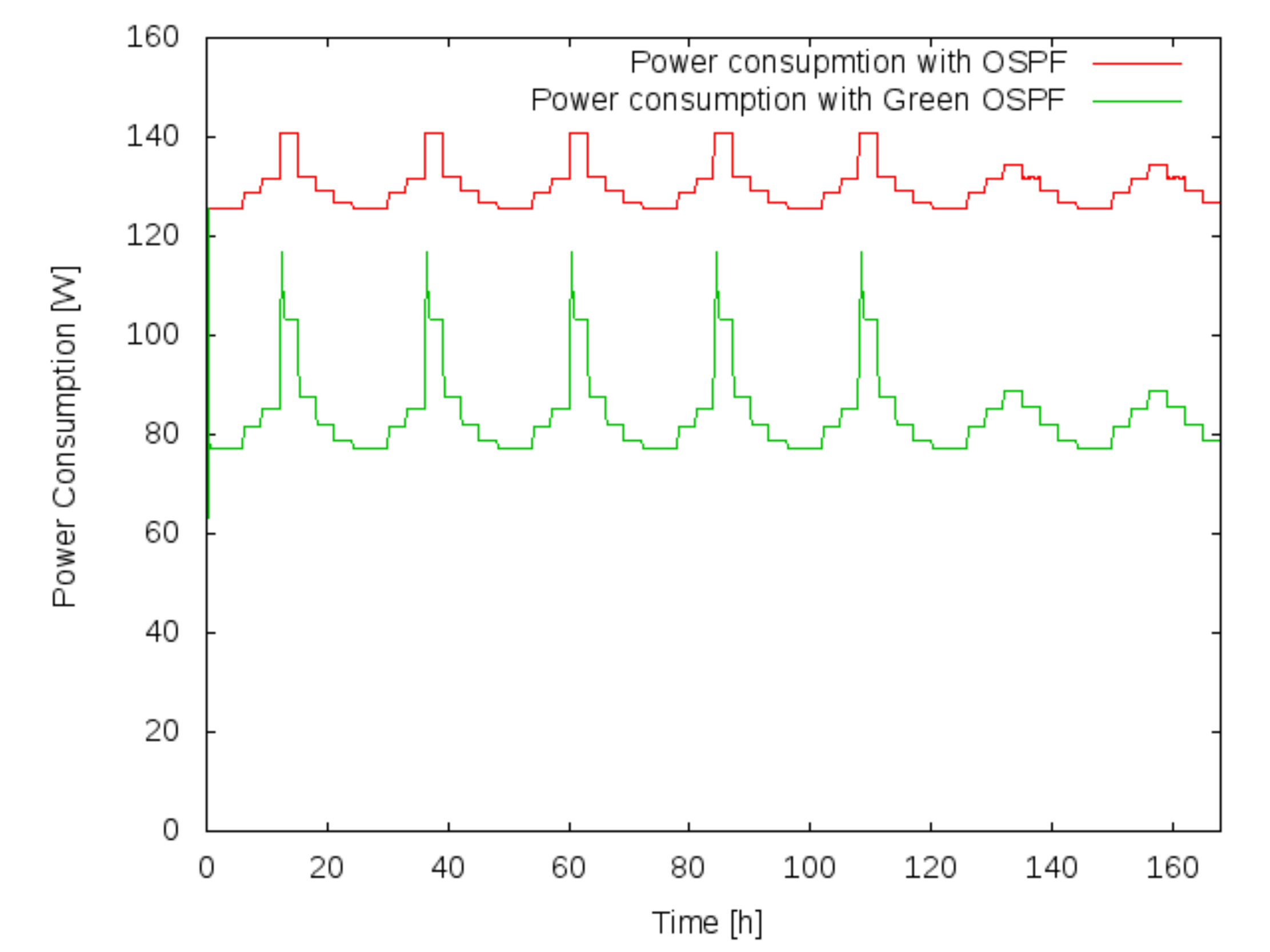}}
  \caption{Testing GoSPF with a weekly UDP traffic profile}
\label{fig:GOSPF_Weekly}
\end{figure}

A weekly UDP traffic is shown in Fig.~\ref{fig:GOSPF_Weekly}(a) where we can note the same trend as in Fig.~\ref{fig:GOSPF_UDP}(a) except for weekends. The number of active links (Fig.~\ref{fig:GOSPF_Weekly}(b)) is set at its minimum during this period.
Once again, we can note that the results that we obtain with this test don't change from the daily test, except for weekends, when energy consumption (see Fig.~\ref{fig:GOSPF_Weekly}(c)) decreases. In particular, energy saving goes from $34,8$\%, with a daily traffic, to $35,5$\% with the weekly traffic test.

%
%

\section{G-OSPF Implementation}
\label{sec:implementation}

We also provided a real implementation of the proposed solution within the popular \emph{Quagga} routing suite~\cite{quagga}. Quagga is an open source package of TCP/IP routing protocols for Unix-like platforms. It supports RIPv1, RIPv2, RIPng, OSPFv2, OSPFv3, BGP-4, and BGP-4+. A command line interface, as well as SNMP (Simple Network Management Protocol) support, make a punctual service configuration and monitoring available. Differently from other similar packages, Quagga relies on a scalable two-tier architecture where a supervisor daemon, namely \emph{zebra}, manages specialized routing daemons, one for each supported protocol. Zebra is responsible for interacting with the operating system's routing tables, thus relieving the protocol daemons from the burden of directly accessing and modifying them. The energy efficient algorithm, which we called \emph{gospf}, is included in Quagga as an extension to the existing OSPF daemon\footnote{Our code is obviously open source and is available for download at the following URL: \texttt{http://wpage.unina.it/spromano/gospf/Gquagga.tar.gz}}.

\subsection{The gospf daemon in Quagga}

This section delves into GOSPF (Green OSPF) implementation details. First of all, we higlight the Quagga modules that are subject to modifications. As described above, any protocol daemon in Quagga makes use of the zebra daemon to modify routing tables. Protocol daemons rely on a management system based on threads and which leverages the select function. Thus, Quagga daemon threads are enqueued and executed in a sequential way. Four types of threads are available, namely \emph{Timed}, \emph{Read}, \emph{Write}, and \emph{Event} threads. \emph{Read} and \emph{Write} threads have a higher priority than \emph{Event} threads, which in turn have higher priority than the \emph{Timed} ones.

To get information like network interfaces status or their actual configuration, a \emph{zclient} structure is initialized in the zebra daemon. Two lists available in the Quagga library, namely \emph{prefix} and \emph{radix} tree, also allow for the storage of important information concerning the daemon protocol execution. Relevant IP address information, for instance, is stored in a prefix element. Routing tables are instead stored in a radix tree list. One further component of the \emph{ospf} daemon is the so called \textit{ospf\_area} data structure, that is a container of information like the link state database and the list of active interfaces. 

The implementation of \emph{gospf} can be seen as an extension to the \emph{ospf} protocol daemon. To the purpose, a new structure is included in the ospf\_area structure to keep track of the Maximum Capacity Spanning Tree (MCST). Each such tree contains a \emph{distance\_table} list, a \emph{mcst\_tree} structure, the \emph{interfaces} radix tree list, a generic \emph{thread} pointer, a further pointer to a specific thread dedicated to MCST computation (\emph{check\_mcst\_thread}) and finally a \emph{reset} notification handler. The distance table is a list of all the networks that are active in a specific area, ordered by their cost. The mcst\_tree contains a table of the routers, with all their interfaces, available in the area. The data provided by these structures enable the daemon to make decisions during the `graft' procedure and then select the interfaces that can be disabled, as well as those that must be kept active. The status of each interface can be set as \texttt{MCST\_TREE}, if part of the MCST, \texttt{MCST\_UNCUT}, if active but out of the MCST, \texttt{MCST\_CUT} when disabled after a `cut' procedure, and \texttt{MCST\_GRAFT} to identify an interface to be kept active during a specified time interval. Additional structures are dedicated to the MCST creation through the execution of Kruskal's algorithm (see Section~\ref{sec:algorithm}). MCST creation is not executed at daemon start-up time; it is rather scheduled after a sufficient number of Link State Advertisement (LSA) messages has been exchanged, which allows a node to gather enough information about its neighbors. Finally, several functions have been added to manage the core of the \emph{gospf} protocol. For instance, a \emph{check\_interface\_bandwidth} function is responsible to verify if the energy consumption level of a specific interface is above a pre-defined threshold, and then proceed with a either a \emph{cut\_interface} or a \emph{restore\_interface} operation. A \emph{check\_interface\_area\_timer} activates a clock to periodically check the status of the interfaces, and some other dedicated functions are responsible for the generation of the newly defined OSPF \texttt{GRAFT} packets.

\subsection{gospf functional testing}

\begin{figure}[h!]
   \centering
       \includegraphics[width=0.85\columnwidth]{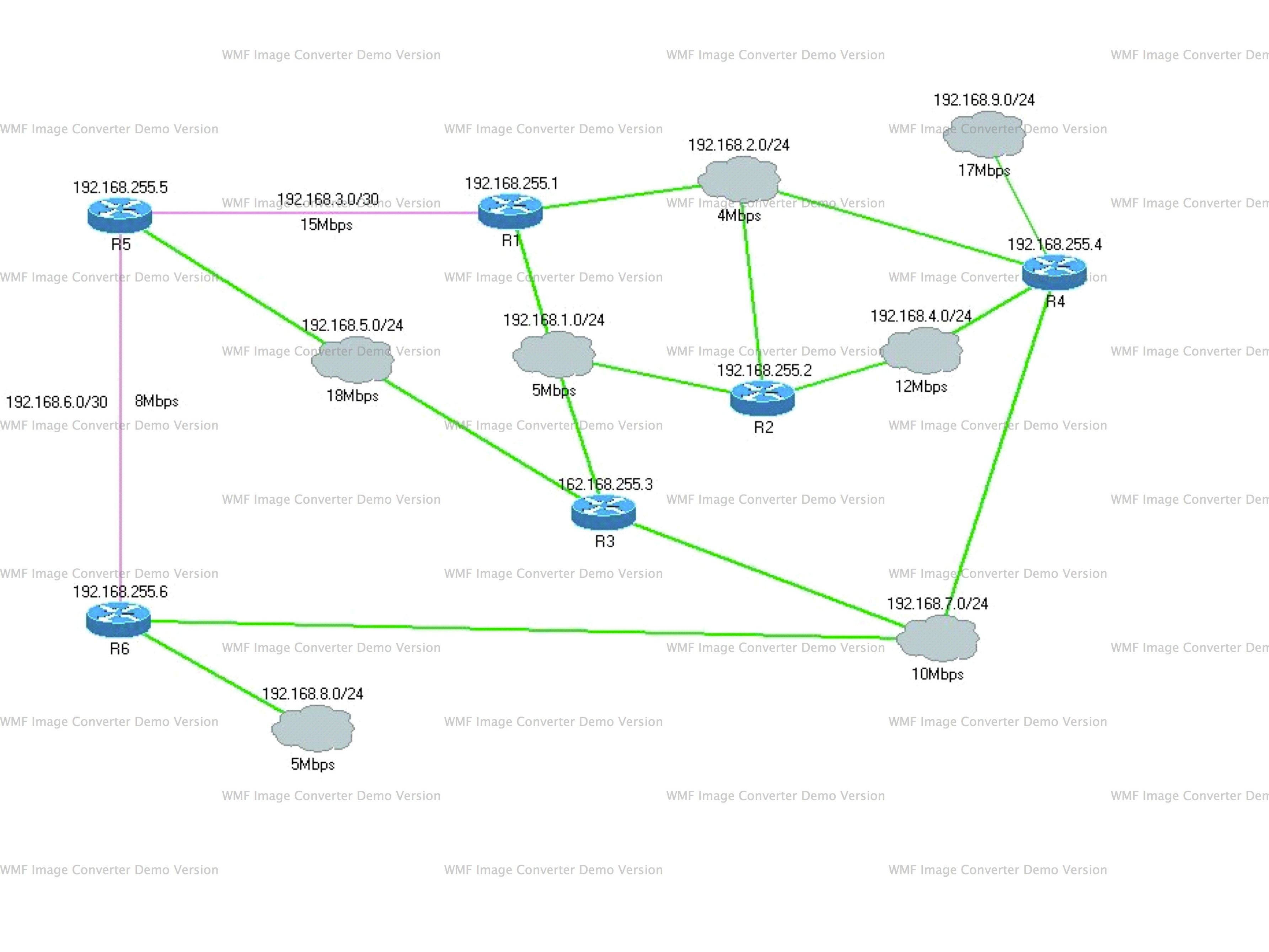}
   \caption{Experimental emulated network topology}
   \label{fig:emuNet}
\end{figure}

The actual \emph{gospf} implemetation has been tested on an emulated network. Several Linux nodes, equipped with our enhanced version of Quagga, have been created through the Qemu\footnote{\texttt{http://www.qemu.org}} emulation environment. The experimental network topology is presented in \figurename\ref{fig:emuNet}. 

\begin{figure}[h!]
   \centering
       \includegraphics[width=0.5\columnwidth]{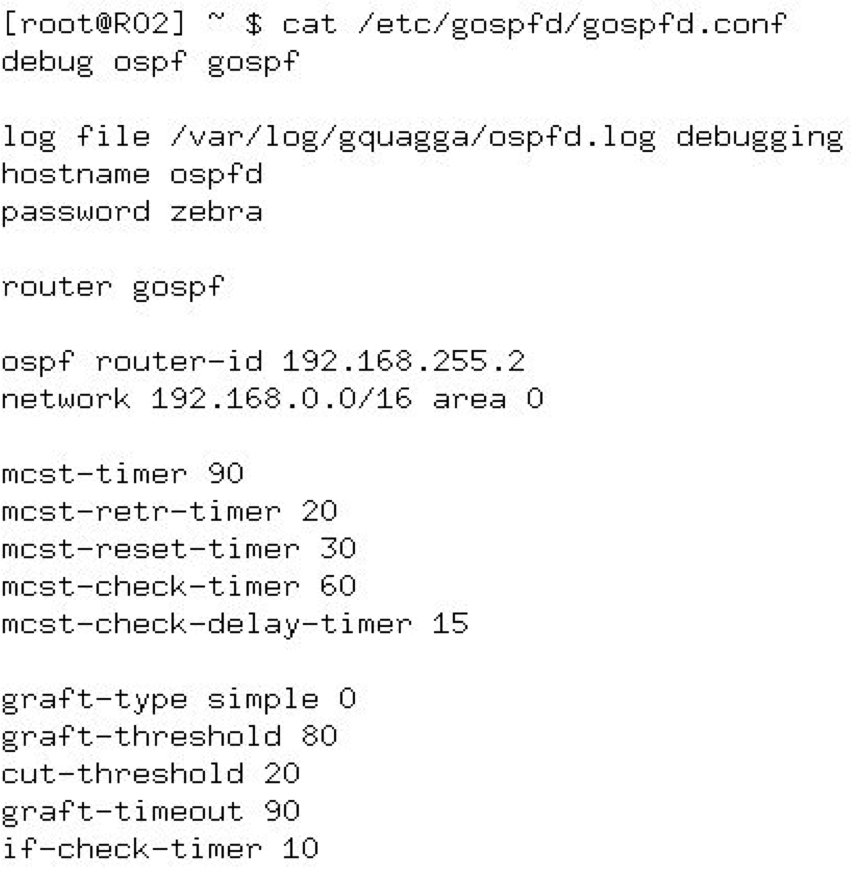}
   \caption{gospfd sample configuration file}
   \label{fig:gospfd_conf}
\end{figure}

Both point-to-point and non broadcast multi access links are included in this topology, all of them based on ethernet connections. We deliberately reproduced artificial competing traffic flows on certain links to verify the correct intervention of the \emph{gospf} algorithm. With respect to router configuration, we report a snapshot of the \emph{gospfd} daemon configuration file associated with Router $2$ in \figurename~\ref{fig:gospfd_conf}. Such a file is indeed almost identical in all gospf-enabled routers.

During the trials, we inspected the log files on routers to check whether the dynamic status of links was compliant to the results expected from the model. 
Each experiment starts with creation of the MCST associated with the network. This is articulated in three steps: (i) creation of a network graph; (ii) MCST computation on the resulting graph; (iii) selection of router interfaces belonging to the MCST.

\begin{figure}[h!]
   \centering
       \includegraphics[width=0.85\columnwidth]{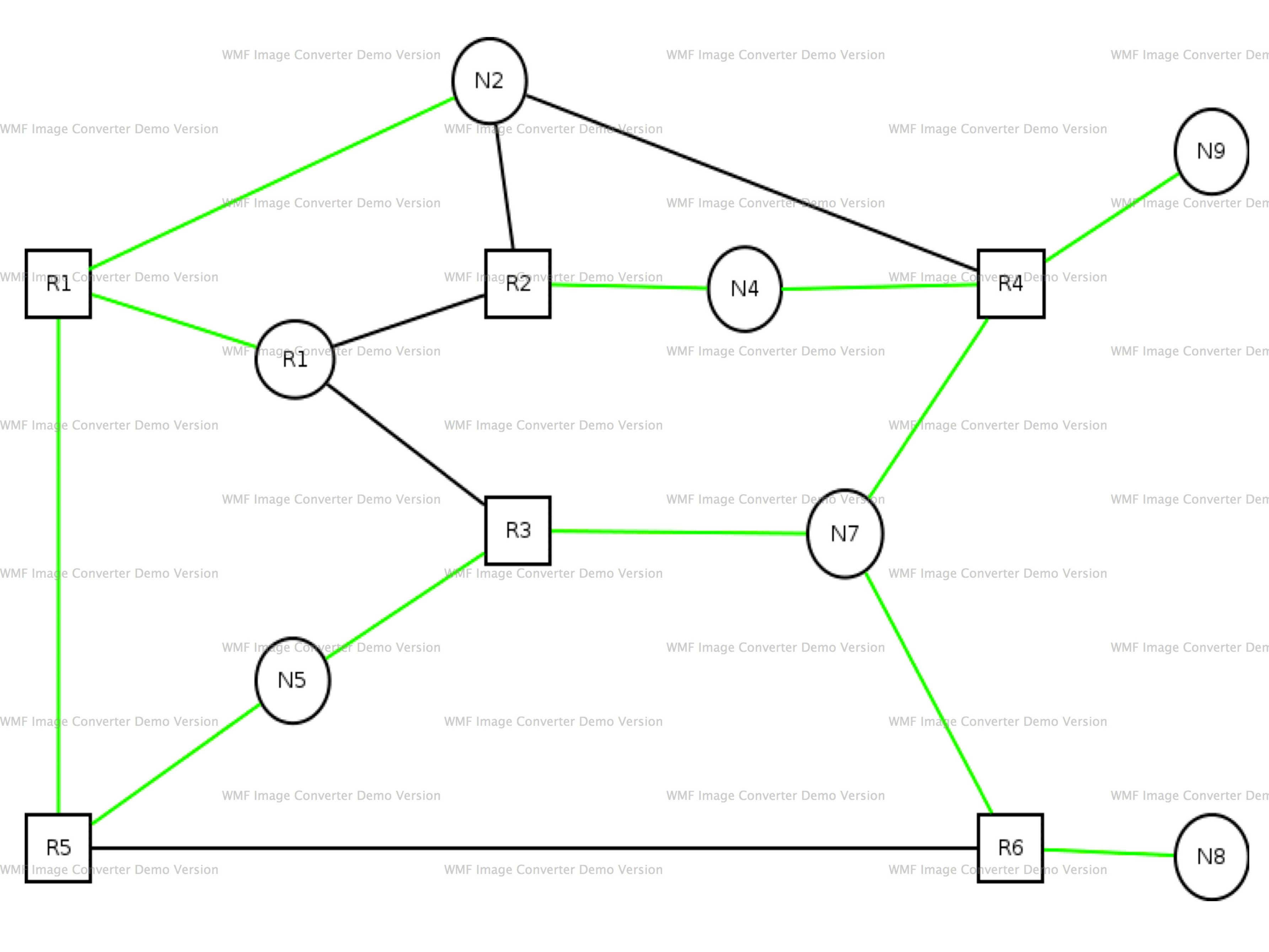}
   \caption{Graph resulting form MCST computation}
   \label{fig:graph}
\end{figure}

The creation of a network graph is achieved through the identification of actual routers and transit networks, labeled respectively as $Rn$ or $Nn$. In this way, MCST computation can guarantee that both routers and nodes located beyond networks are reachable. For the experiment in question, the resulting graph is reported in \figurename\ref{fig:graph}. The computed minimum spanning tree (which, as mentioned in Section~\ref{sec:algorithm}, is in our case a Maximum Capacity Spanning Tree) is represented in green in the picture.

\begin{figure}[h!]
   \centering
       \includegraphics[width=0.8\columnwidth]{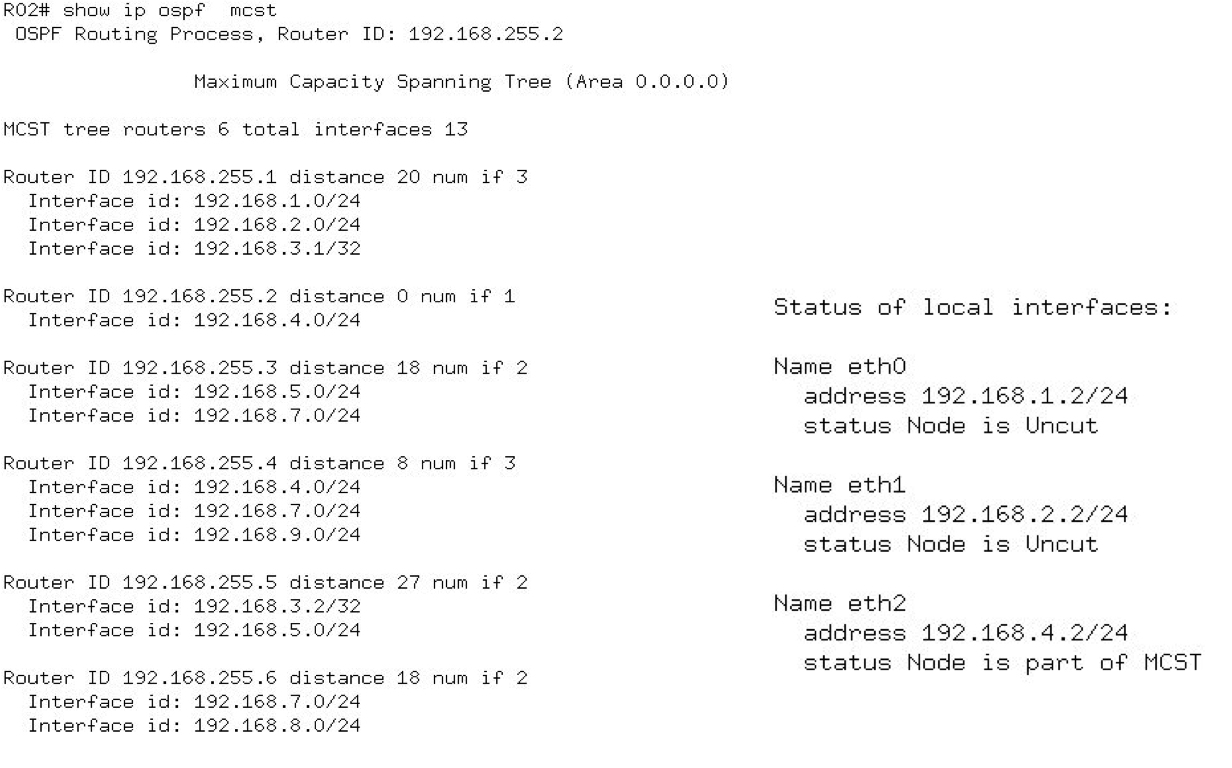}
   \caption{Dumping GOSPF status on R2's console}
   \label{fig:R2_dump}
\end{figure}

The last step concerns the selection of the relevant interfaces to be assigned to the MCST. To the purpose, we tested the behavior of the \emph{gospfd} when some traffic is generated in the network. From the picture, it is clear that all of the interfaces of router $R1$ belong to the MCST and hence can never be cut. On the other hand, router $R2$ owns three interfaces, just one of which cannot be cut since it is crucial for keeping nodes reachability across the MCST. More precisely, while setting up the experiment, the mentioned interfaces were configured as follows: (i) eth0 (192.168.1.2/24) has a 5Mbits/s capacity and can be cut off; (ii) eth1 (192.168.2.2/24) has a 4Mbits/s capacity and can be cut off; (iii) eth2 (192.168.4.2/24) has a 12Mbits/s capacity, it is part of the MCST and thus cannot be cut off. All of the above considerations are actually confirmed by \figurename~\ref{fig:R2_dump}, which reports a dump of the execution of a ``\texttt{show ip ospf mcst}'' command on R2's console.

\subsubsection{Cutting}

\begin{figure}[h]
  \centering
  \subfigure[Interfaces configuration on R2]{\includegraphics[width=0.55\textwidth]{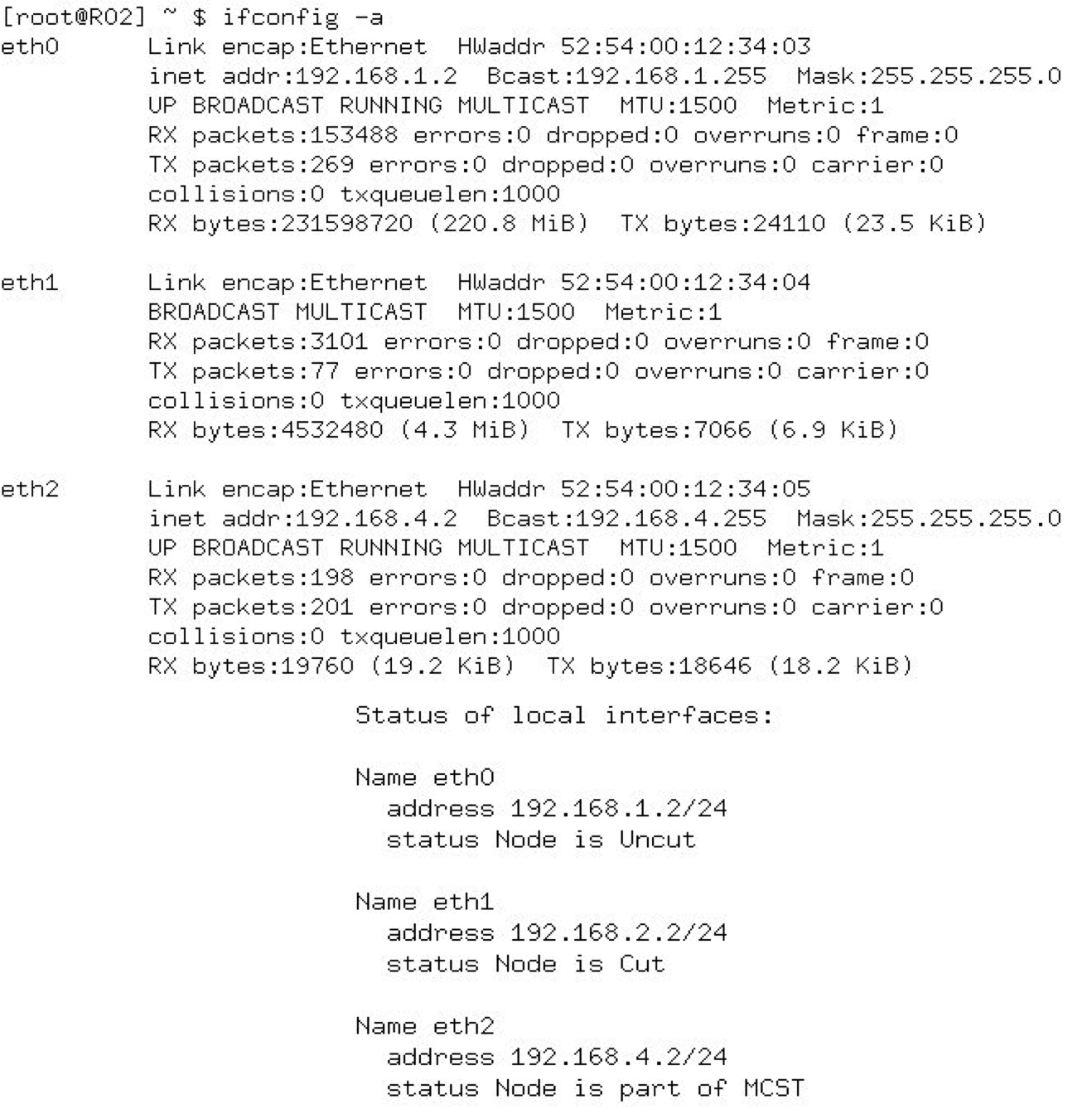}}
  \subfigure[LSA database on R1]{\includegraphics[width=0.42\textwidth]{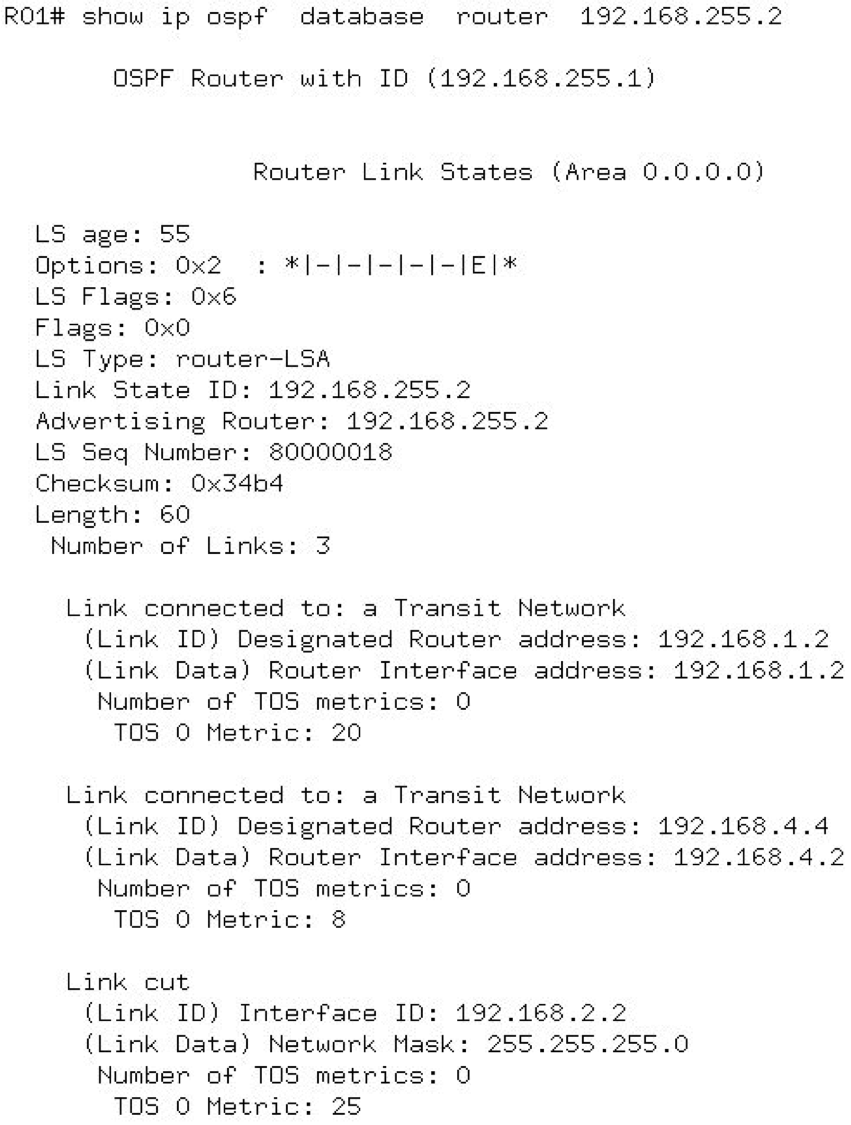}}
  \caption{GOSPF cutting effects on routers' configurations}
\label{fig:Cutting}
\end{figure}

In the above scenario, if a 3 Mbit/s constant bit rate traffic is generated on interface eth0, and a lighter 500 Kbit/s flow crosses interface eth1, the gospf protocol makes the decision to cut eth1 because its utilization is under the threshold, while keeping eth0 active because its utilization level falls in between its normal usage interval. Hence, a \texttt{Router\_LSA} is sent to the other routers to notify them about the updates. \figurename~\ref{fig:Cutting}(a) shows that eth1 has been deactivated on R2 (there is no IP address assigned to it and it is flagged as `Cut') due to the cutting procedure operated by \emph{gospfd}. Similarly, \figurename~\ref{fig:Cutting}(b) offers a view of the LSA database on router R1, which clearly demonstrates how \emph{ospfd} has spread the cutting news to the other routers in the network (`Link cut' section on the bottom of the picture).

\subsubsection{Grafting}

\begin{figure}[h!]
   \centering
       \includegraphics[width=0.85\columnwidth]{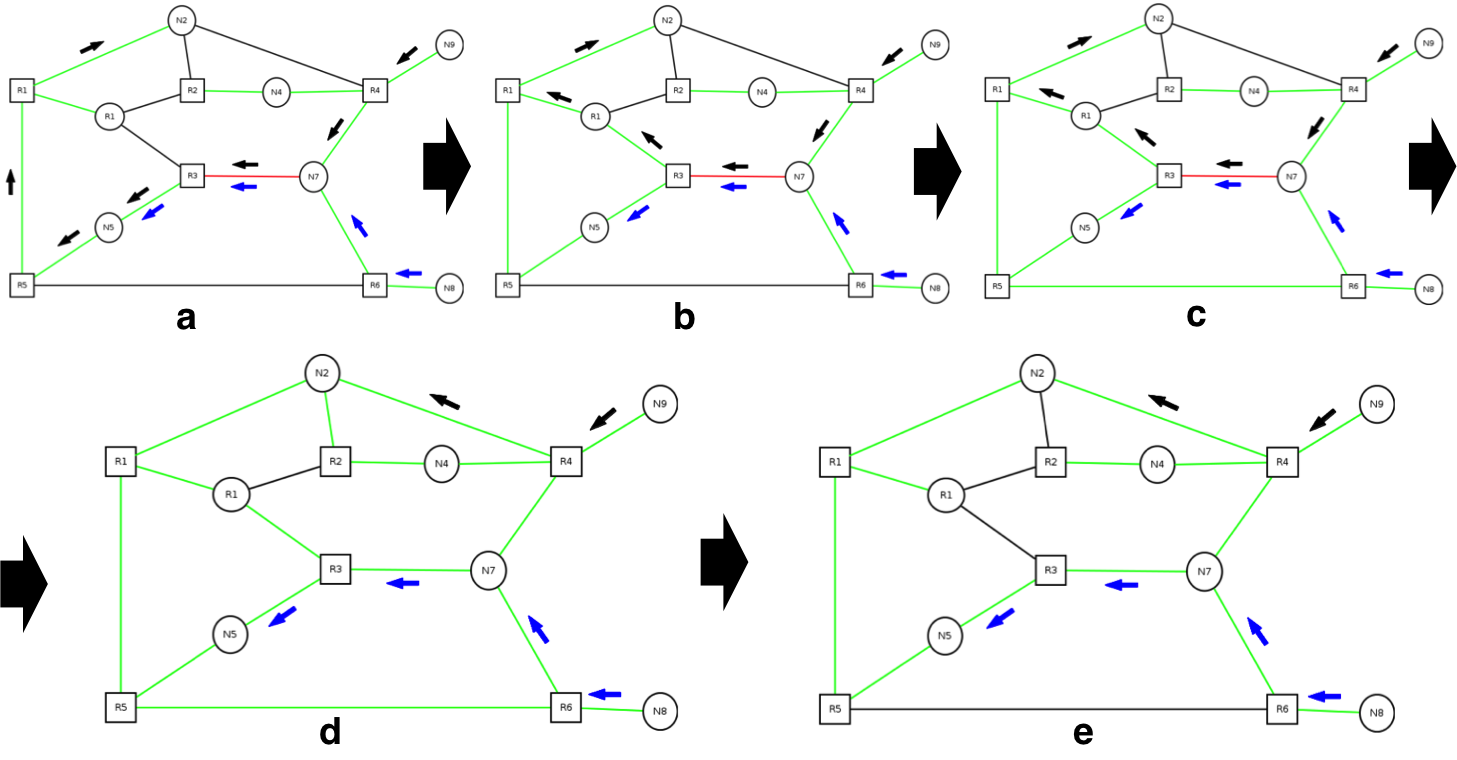}
   \caption{G-OSPF reaction to congestion}
   \label{fig:gospf}
\end{figure}


We then evaluate the opposite situation, that is \emph{gospfd}'s reaction to a congestion state. We start with the assumption that all of the interfaces not pertaining to the MCST are disabled. Now, two new traffic flows are injected into the network: the former is a 3Mbit/s constant bit rate from host H9 belonging to network 192.168.9.240/24 (network N9) to host H2 belonging to network 192.168.2.240/24 (network N2); the latter is a 6Mbit/s constant bit rate from host H7 on network 192.168.7.240/24 (network N7) to host H5 on network 192.168.5.240/24 (network N5). In the depicted scenario, network N7 will be crossed by a 9Mbit/s aggregated traffic which will cause congestion on the link N7-R3 (whose capacity is 10Mbit/s) highlighted in red in \figurename~\ref{fig:gospf}(a).

According to the GOSPF protocol, this causes a \texttt{GRAFT} request. The first action is thus to awake all of the local interfaces, which in this case produces the reactivation of the link between router R3 and router R1 (\figurename~\ref{fig:gospf}(b)). As this does not solve the congestion situation, a new \texttt{GRAFT} request is solicited, this time pointed towards other routers in the network (since R3 has no more interfaces in the `Cut' state). Rather than flooding this request, R3 first checks its local information in search of routers or networks, ordered by distance, having interfaces in a `Cut' state. As network 192.168.6.0/30 has interfaces forced to be down, the \texttt{GRAFT} request is sent towards routers R5 and R6 and asks for the activation of their disabled interfaces (\figurename~\ref{fig:gospf}(c)). Again, this countermeasure turns out to be not effective, and a further \texttt{GRAFT} request is sent towards routers R2 and R4 to demand for the activation, respectively, of the interfaces 192.168.2.2/24 and 192.168.2.4 (\figurename~\ref{fig:gospf}(d)). This action finally solves the congestion problem. However, GOSPF executes a last step to cut off the unnecessary interfaces (i.e., those crossed by a low-density traffic flow), leading to the final topology depicted in \figurename~\ref{fig:gospf}(e).

\subsubsection{Resetting GOSPF}

\begin{figure}[h]
  \centering
  \subfigure[Original situation]{\includegraphics[width=0.46\textwidth]{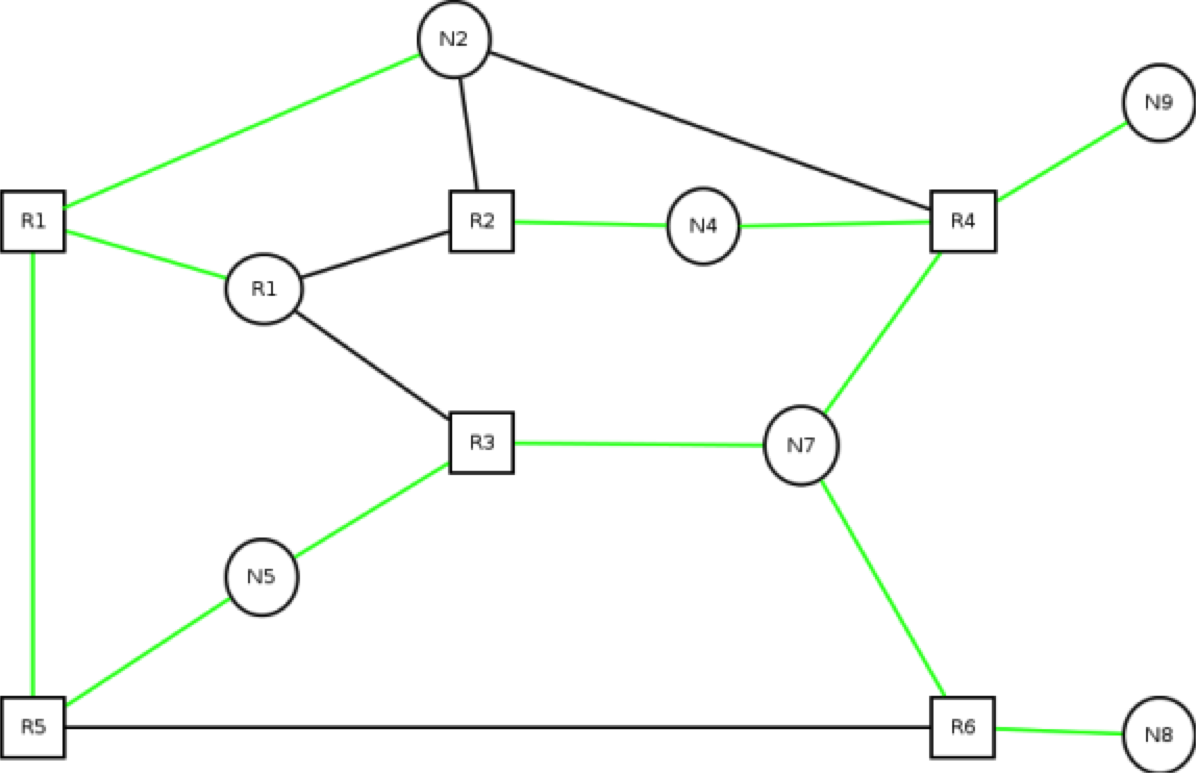}}
  \subfigure[Link failure]{\includegraphics[width=0.46\textwidth]{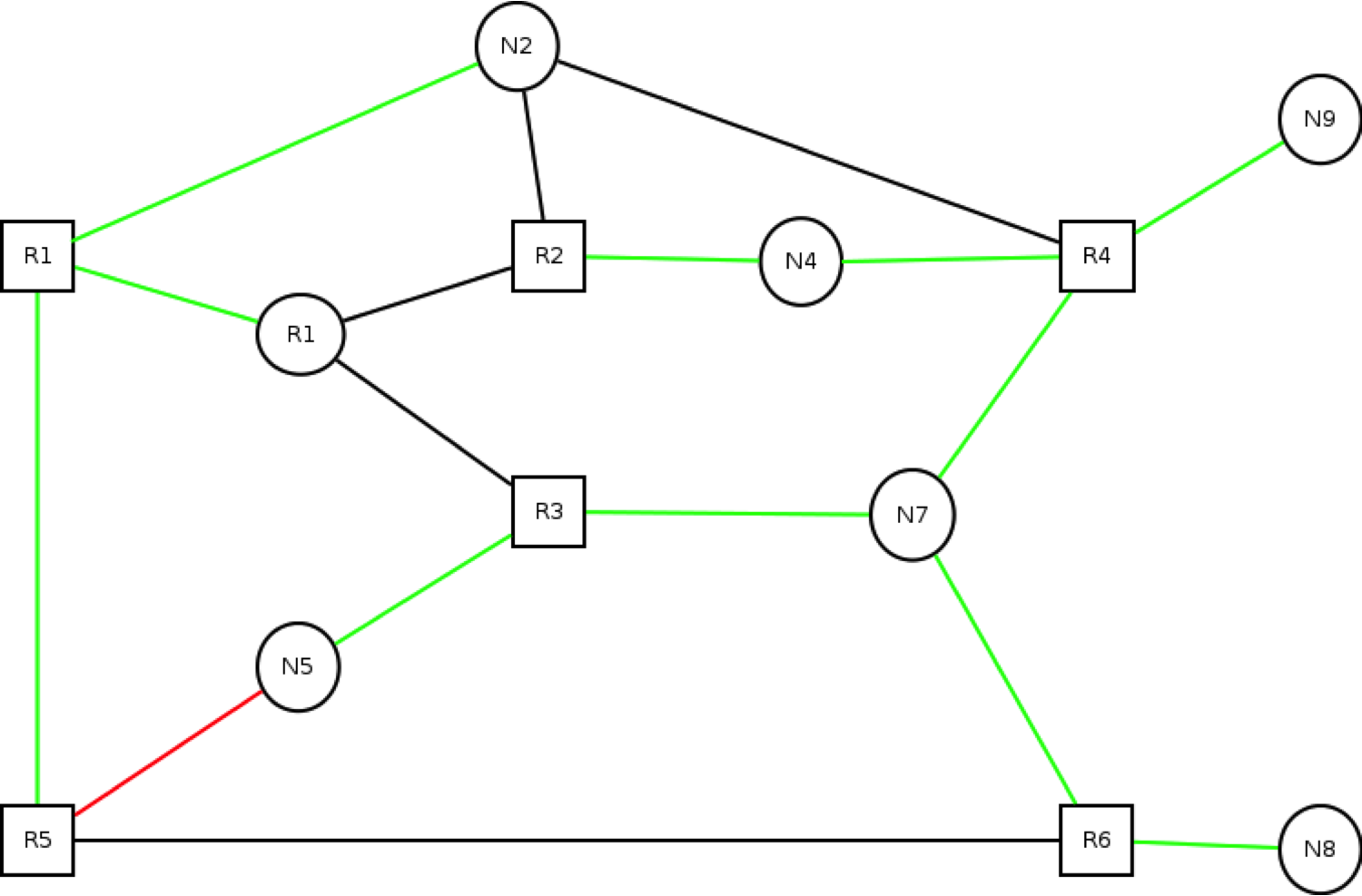}}
  \subfigure[Interface reactivation]{\includegraphics[width=0.46\textwidth]{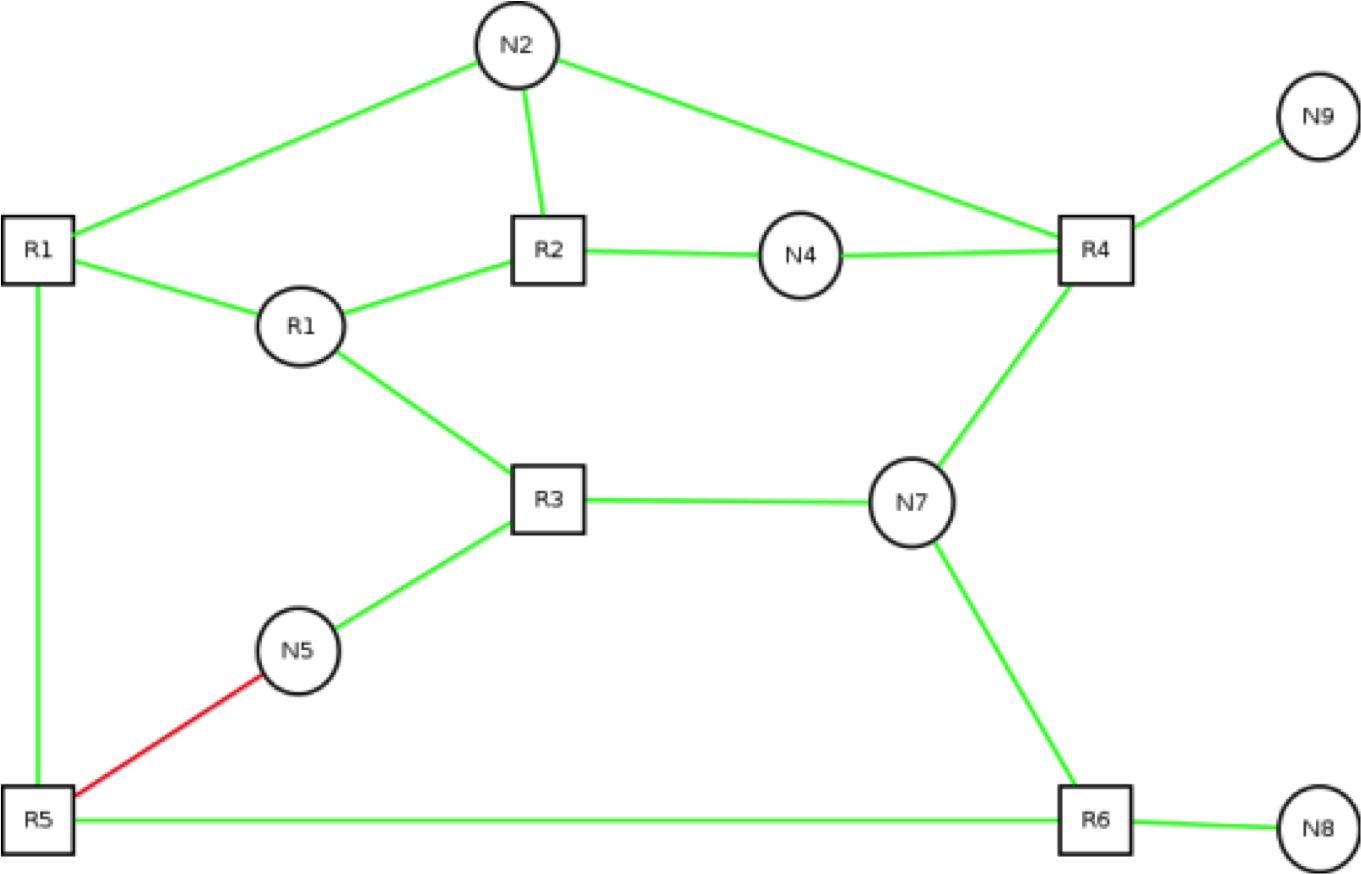}}    
  \subfigure[New MCST]{\includegraphics[width=0.46\textwidth]{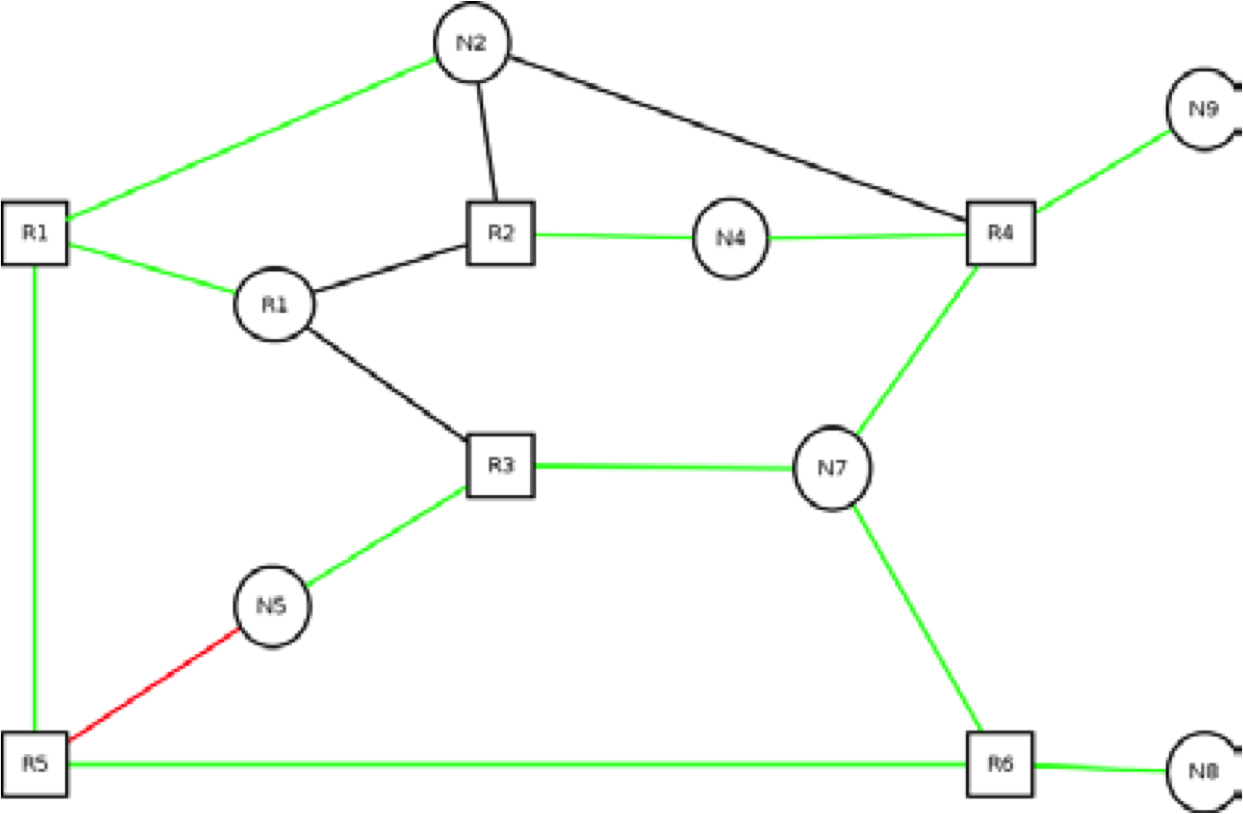}}
  \caption{GOSPF dealing with a link failure through a \texttt{RESET} operation}
\label{fig:Reset}
\end{figure}

We finally illustrate how GOSPF is capable to dynamically react to unpredicted topology modifications (affecting the MCST) due to link failures. Starting from a steady-state situation as the one depicted in \figurename~\ref{fig:Reset}(a), let us assume that link R5-N5 goes down (which can be easily achieved by forcibly shutting down one of the interfaces on R5). In this situation (\figurename~\ref{fig:Reset}(b)), as soon as a router initiates the periodically scheduled MCST integrity check task it identifies an error, hence triggering the \emph{reset} procedure. 

%

As a result of the above operation, all network interfaces are activated, hence bringing to the new graph illustrated (in green) in \figurename~\ref{fig:Reset}(c). After a while (as determined by the \texttt{mcst-reset-timer} parameter of GOSPF), a brand new MCST is computed and configured on the network, as shown in \figurename~\ref{fig:Reset}(d).

\section{Security considerations}
\label{security}

Given that GOSPF is an interior routing protocol, one might question the need for securing their routing environment given that their infrastructure is entirely contained within a single organization and hence somehow isolated from the outside world. Actually, besides providing an additional layer of defense (which is in-line with the so-called ``defense in depth'' approach), securing GOSPF routing enables protection against internal threats, which is something going beyond what firewalls, Intrusion Detection Systems and Virtual Private Networks can do. After all, recent statistics mention up to 70\% of information security threats come from within a network's perimeters. 
As any other routing protocol, GOSPF might indeed be the target of attacks form malicious users aimed at compromising the correct operation of the network. As an example, a specific attack vector might rely on corrupting certain nodes in order to force them to advertise fake underloaded links, hence inducing GOSPF to cut them and force a path across enemy-controlled links or nodes. 

We also remark that securing GOSPF networks will protect them from not only malicious attacks, but also accidental misconfigurations. The friendly nature of the underlying OSPF protocol dictates that, by default, any router with coordinated configuration parameters (network mask, hello interval, dead interval, etc.) can participate in a given OSPF network. Because of this default behavior, any number of accidental factors (misconfigurations, test setups, etc.) can adversely affect routing in an OSPF-based environment.

The first step to take in securing GOSPF is to configure all participating devices as non-broadcast devices. In non-broadcast, or directed, mode, OSPF devices need to be explicitly configured to communicate with valid OSPF neighbors. This provides a basic layer of security against misconfiguration, because valid OSPF devices will only communicate with the OSPF devices they have been configured to interoperate with. On the other hand, in a broadcast OSPF environment any OSPF device with the correct configuration parameters for the network will be able to participate in OSPF routing. On most OSPF-enabled routers, interfaces use broadcast OSPF by default; though ``directed OSPF'' mode can be turned on via explicit router commands.

As a second step, to increase the security of a GOSPF environment, OSPF authentication can be enabled. Indeed, OSPF authentication can be either `none', `simple' or `MD5'. With simple authentication, the password goes in clear-text over the network. Thus, anyone capable to sniff traffic on the OSPF network segment would be able to retrieve the OSPF password, and the attacker would be one step closer to compromising the target OSPF environment. With MD5 authentication, the key does not pass over the network. MD5 is a message-digest algorithm specified in RFC1321 and can be considered the most secure OSPF authentication mode.

With the steps suggested above, a vulnerable dynamic GOSPF-based environment gets converted over to directed and MD5 authenticated GOSPF. Of course, this does not mean that the routing environment has now been turned into a 100\% secure one, but its stealthiness level has definitely been increased. One thing to consider regarding the topics discussed thus far, is that the MD5 authentication process does not provide encryption of routing data. Instead, it verifies the sending and receiving parties. Thus, an attacker with a sniffer on the wire would still be able to decode routing update packets. Nonetheless, injecting false packets and compromising the environment will just be a lot more difficult. Also, issues can arise when keys are modified. If one follows best security practices, GOSPF authentication keys will get regularly replaced. 

We finally remark that a further level of security can be added by leveraging some form of router trustworthiness information within the GOSPF network. Namely, if one associates with each router a properly computed (and dynamically updated) trustworthiness level, it becomes possible to not rely on updates sent by unreliable neighbors, so to avoid malicious modifications of the Link State Database causing unwanted cuts of links, as well as switching off of some nodes. With respect to this specific field of investigation, our previous work in~\cite{Refacing} describes a generic approach to the computation of the level of trustworthiness of a set of distributed yet interoperating nodes. The work in~\cite{Globecom2008} instead focuses on routing, even though in a wireless networking scenario and in the presence of a distance-vector based routing protocol.

\section{Conclusions}
\label{sec:conclusions}

In this paper we presented an OSPF extension aimed at minimizing network power consumption while at the same time keeping both connectivity and a satisfactory level of Quality of Service. The solution we devised dynamically adapts the network topology by switching off underutilized links of the network. The distributed algorithm we propose is based on a `graft and prune' strategy and has been formally specified with an analytical model taking into account, among other requirements, traffic demand dynamics. A thorough study of the performance of our proposal, in terms of both energy efficiency and induced overhead, has been conducted through simulations which closely reproduce a real-world scenario associated with the Italian research network topology and supported traffic patterns. Furthermore, a real-world implementation of the proposed GOSPF protocol has been described and an extensive overview of typical operational scenarios has been provided.

Among the future directions of work we definitely see the following main points. First, a detailed energy consumption model, reliably mimicking the energy profiles of currently deployed network routers, has to be embedded in our power consumption computations. Second, a dynamically adjustable way of setting the two main thresholds involved in the execution of our algorithm (associated, respectively, with the triggering of grafting and pruning operations) has to be defined. We are already working on this task, which clearly requires to strike a balance between promptness of reaction (i.e., adaptability to changes in the supported traffic profile) and stability (i.e., capability to avoid oscillations and route flaps due to the high frequency of adaptation to traffic changes). 


\section*{References}









\end{document}